\documentclass{article}

\usepackage{PRIMEarxiv}

\usepackage[utf8]{inputenc} % allow utf-8 input
\usepackage[T1]{fontenc}    % use 8-bit T1 fonts
\usepackage{hyperref}       % hyperlinks
\usepackage{url}            % simple URL typesetting
\usepackage{booktabs}       % professional-quality tables
\usepackage{amsfonts}       % blackboard math symbols
\usepackage{nicefrac}       % compact symbols for 1/2, etc.
\usepackage{microtype}      % microtypography
\usepackage{lipsum}
\usepackage{fancyhdr}       % header
\usepackage{graphicx}       % graphics
\graphicspath{{media/}}     % organize your images and other figures under media/ folder
\usepackage{amsfonts}
\usepackage[utf8]{inputenc}
\usepackage[T1]{fontenc}
\usepackage{amsmath}
\usepackage{bm}
\usepackage{multirow}
\usepackage[draft]{pgf}
\usepackage{hyperref}
\usepackage{makecell}
\usepackage{float}
\usepackage{lineno}
\usepackage{wrapfig}
\usepackage{amssymb}
\usepackage{caption}
\usepackage{subcaption}

\title{\textsc{DeFault}: DEep-learning-based FAULT Delineation Using the IBDP Passive Seismic Data at the Decatur CO2 Storage Site
}

\author{
  Hanchen~Wang \\
  Theoretical Division \\
  Los Alamos National Laboratory\\
  Los Alamos, NM 87545, USA \\
  University of North Carolina at Chapel Hill \\
  Chapel Hill, NC, USA \\
  \texttt{hanchen.wang@lanl.gov} \\
   \And
Yinpeng~Chen \\
Google DeepMind \\
WA, USA \\
\texttt{yinpengc@google.com} \\
\And
Tariq~Alkhalifah \\
King Abdullah University of Science and Technology \\
Thuwal, KSA \\
\texttt{tariq.alkhalifah@kaust.edu.sa} \\
\And
Ting~Chen \\
Los Alamos National Laboratory\\
Los Alamos, NM 87545, USA \\
\texttt{tchen@lanl.gov} \\
\And
Youzuo~Lin \\
University of North Carolina at Chapel Hill \\
Chapel Hill, NC, USA \\
\texttt{yzlin@unc.edu}
\And
David~Alumbaugh \\
Lawrence Berkeley National Laboratory \\
Berkeley, CA, USA \\
\texttt{dlalumbaugh@lbl.gov}
}

\begin{document}
\maketitle
% \begin{keypoints}
% \item Faults and fractures activated by carbon storage can be monitored by passive seismicity. 
% \item $\mathbf{\emph{DeFault}}$ algorithm enables an automatic process for accurate and efficient passive seismic event locating and clustering.
% %\item 
% \end{keypoints}

\begin{abstract}
The carbon capture, utilization, and storage (CCUS) framework is an essential component in reducing greenhouse gas emissions, with its success hinging on the comprehensive knowledge of subsurface geology and geomechanics. Passive seismic event relocation and fault detection offer vital insights into subsurface structures and the ability to monitor fluid migration pathways. Accurate identification and localization of seismic events, however, face significant challenges, including the necessity for high-quality seismic data and advanced computational methods. To address these challenges, we introduce a novel deep learning method, $\mathbf{\emph{DeFault}}$, specifically designed for passive seismic source relocation and fault delineating for passive seismic monitoring projects. By leveraging data domain-adaptation, $\mathbf{\emph{DeFault}}$ allows us to train a neural network with labeled synthetic data and apply it directly to field data. Using $\mathbf{\emph{DeFault}}$, the passive seismic sources are automatically clustered based on their recording time and spatial locations, and subsequently, faults and fractures are delineated accordingly. We demonstrate the efficacy of $\mathbf{\emph{DeFault}}$ on a field case study involving CO$_2$ injection related microseismic data from Decatur, Illinois area. Our approach accurately and efficiently relocated passive seismic events, identified faults and could aid in potential damage induced by seismicity. Our results highlight the potential of $\mathbf{\emph{DeFault}}$ as a valuable tool for passive seismic monitoring, emphasizing its role in ensuring CCUS project safety. This research bolsters the understanding of subsurface characterization in CCUS, illustrating machine learning's capacity to refine these methods. Ultimately, our work has significant implications for CCUS technology deployment, an essential strategy in combating climate change.
\end{abstract}

% \vspace{-4mm}

% keywords can be removed
% \keywords{Joint diffusion model \and Wave equation  \and Full waveform inversion}

\section{Introduction}\label{sec:introduction}

Carbon capture, utilization, and storage (CCUS) has emerged as a pivotal framework in the global effort to combat climate change. By targeting CO$_2$ emissions from industrial and energy-related sources, CCUS offers a promising avenue to significantly reduce the carbon footprint of human activities \cite{pacala2004stabilization}. A cornerstone of the success of CCUS projects lies in the meticulous monitoring and characterization of subsurface geological structures. Among these, faults are of particular concern, as they can jeopardize the long-term containment of stored CO$_2$ \cite{white2016assessing}. Thus, the detection, analysis, and understanding of these faults are paramount to ensure the safety, efficiency, and public acceptance of CCUS technologies.

The Illinois Basin–Decatur Project (IBDP) stands as a testament to the potential of CCUS. Located at the Archer Daniels Midland (ADM) Company’s ethanol processing plant in Decatur, Illinois, the IBDP is a large-scale demonstration project for integrated carbon capture and geological storage in a saline reservoir. With nearly 1 million tonnes of CO$_2$ injected from November 2011 to November 2014, and a subsequent second injection well (CCS2) introduced in 2017, the project has been at the forefront of CCUS implementation, providing invaluable data and insights \cite{bauer2019,finley2014}.
\begin{figure*}[!ht]
\centering
\includegraphics[trim={0cm 2cm 0cm 4cm},clip,width=1\columnwidth]{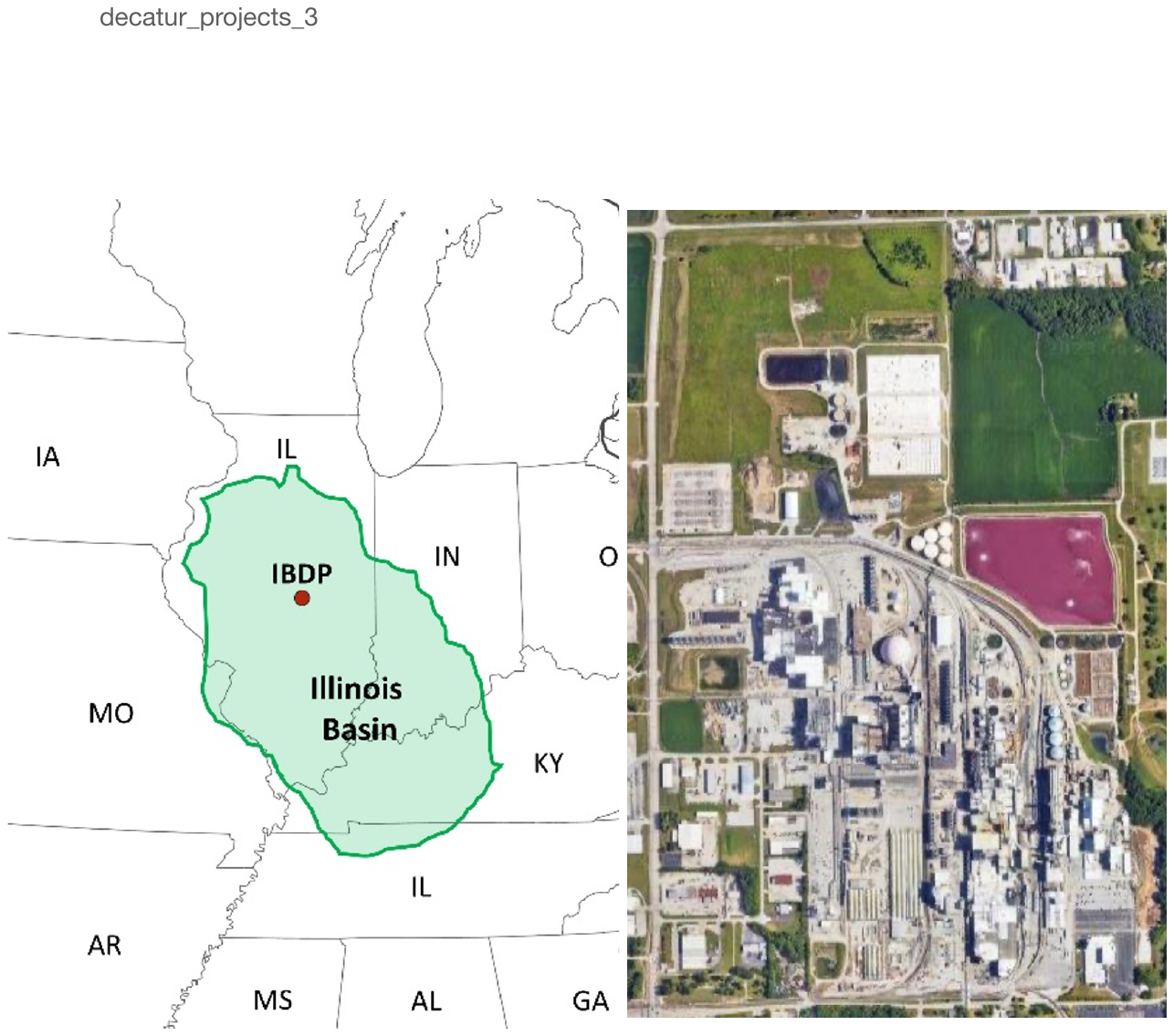}
  \caption{Illinois basin regional map (left), and Decatur CO$_2$ site map view (right).}
\label{fig:decatur1}
\end{figure*}
The Decatur site, shown in Figure~\ref{fig:decatur1}, has been a hub of scientific inquiry. Comprehensive passive seismic monitoring data from the site offers the opportunity for a detailed look into the subsurface structures, serving as a fertile ground for the development, testing, and validation of novel methodologies. With the global urgency to address climate change and the potential of CCUS as a solution, projects like the Decatur initiative are of paramount importance. They serve as real-world implementations of CCUS technologies and as living laboratories, fostering innovation, research, and the development of best practices in the field \cite{vasco2008reservoir,JIANG2021110521,TAPIA20181,HASAN20152}.

Conventionally, seismic methods require a preliminary absolute location of seismic events, usually obtained via a catalog or hypoinverse approach, before applying techniques like Hypo-DD~\cite{waldhauser2000double} and Tomo-DD~\cite{zhang2003double}. These methods, which refine the accuracy of seismic event locations based on these initial estimates, are crucial for a deeper understanding of subsurface structures and offer valuable insights into fault geometry, activity, and potential risks. In the context of the Decatur project, these methods have been employed to gain a deeper understanding of the subsurface behavior of the injected CO$_2$ and the surrounding geological formations. Their results have significantly contributed to our understanding of the key processes and challenges associated with CCUS, offering a comprehensive view of the subsurface dynamics at play. 

On the other hand, waveform inversion techniques have steadily gained traction in the realm of passive seismic monitoring, offering a nuanced perspective of subsurface dynamics. Methods such as full waveform inversion (FWI) \cite{kamei2014passive,sun2016full,wang2018microseismic,8883032,wang2020regularized} and time-reversal migration (TRM) \cite{artman2010source,nakata2016reverse,wang2017time,lyu2020iterative,chen2015directly} stand out, elevating the precision of interpreting seismic signals and enhancing our grasp of underlying geological structures. These techniques excel in extracting intricate details from seismic signals, promising unprecedented clarity into the underlying geological formations. However, their
computational cost has limited their application as near-real time monitoring tools in projects like Decatur and other CCUS ventures.

Recent advances in machine learning (ML) have opened up new possibilities for seismic data analysis and interpretation. A variety of ML techniques, including supervised and unsupervised learning, have been applied to various aspects of subsurface characterization~\cite{bergen2019preface}, including fault detection \cite{wang2018successful,xiong2018seismic,zheng2019applications,hu2020seismic}. Previous works, such as \cite{ma2022small,ANIKIEV2023104371}, have shown the applications of ML-based methods focused primarily on event detection and location. None of them considers the direct delineation of faults from raw seismic monitoring data.

\begin{figure*}[!ht]
\centering
\includegraphics[trim={0cm 0cm 0cm 2cm},clip,width=1\columnwidth]{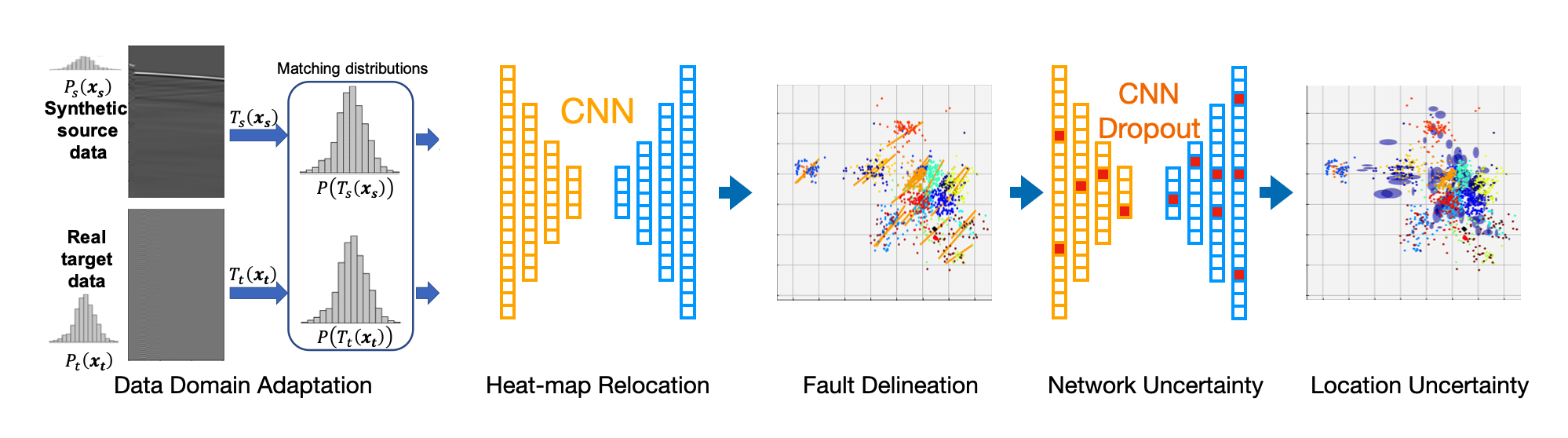}
  \caption{Scratch workflow of the $\mathbf{\emph{DeFault}}$ algorithm.}
\label{fig:teaser}
\end{figure*}

In this study, we aim to develop and validate an ML-based fault detection method, $\mathbf{\emph{DeFault}}$, using passive seismic monitoring data from the Decatur CO$_2$ site. By utilizing cutting-edge ML techniques, shown as the scratch workflow in Figure~\ref{fig:teaser}, we seek to improve the efficiency and accuracy of fault detection and analysis, thereby contributing to the safety and success of the IBDP CCUS projects.
Our objectives in this study are as follows:
\begin{enumerate}
\item Develop an ML-based fault detection method, $\mathbf{\emph{DeFault}}$, by incorporating passive seismic monitoring data from the Decatur CO$_2$ site. This involves training ML models on labeled datasets and validating their performance on unseen data.
\item Develop a domain adaptation method that allows us to generate realistic synthetic training sets sharing the same data distribution to the real monitoring seismic data.
\item Enhance the neural network models' generalization capabilities by aligning the distributions of synthetic and field data. This aims to improve model performance across diverse operations by ensuring the models can effectively interface with field data.
\item Compare the performance of $\mathbf{\emph{DeFault}}$ with the catalog of a modified double-difference relocation methods~\cite{dando2021relocating} in terms of accuracy, efficiency, and applicability to complex geological settings. This involves analyzing the relocation results, fault geometries, and associated uncertainties for the relocation.

\end{enumerate}
To achieve these objectives, we will first compile and preprocess the necessary data from the Decatur CO$_2$ site and other sources, including passive seismic monitoring data, and any available geological and geophysical information. We will then train and validate our ML models on these datasets, using the MLReal technique outlined in \cite{alkhalifah2022mlreal}. Finally, we will assess the uncertainty of $\mathbf{\emph{DeFault}}$ event relocation results. This will involve adapting a dropout-enabled neural network architecture to multiple independent field data tests. By achieving these objectives, we aim to contribute to the ongoing development and implementation of effective fault detection and analysis methods for CCUS projects.

\section{Previous Studies of the Decatur Project}

The Illinois Basin-Decatur Project (IBDP) stands as a testament to the feasibility of carbon capture and geologic storage, having successfully sequestered nearly 1.1 million tons of CO$_2$ within the Mt. Simon Sandstone unit at a depth of 2.14 km. \cite{bauer2016overview} provided an in-depth examination of the project's thorough pre-injection site investigations, detailed monitoring throughout the injection phase, and dedicated post-injection surveillance. A significant aspect of their investigation was the microseismic activity triggered by CO$_2$ injection, highlighting the reservoir's geomechanical responses. While their study contributed valuable data on the possible microseismic reactions to CO$_2$ injection, the interpretation of induced seismicity as a definitive hazard remains a subject of ongoing research and debate. Nonetheless, the insights gained from IBDP have been influential, notably impacting the subsequent completion strategies for the CCS2 well. This acknowledgment of the complexities involved in assessing microseismic responses and their implications provides a foundation for refining the safety and efficiency protocols in future CO$_2$ storage projects.

The study of \cite{kaven2014seismic} explored the seismic monitoring activities at the Decatur, IL, CO$_2$ sequestration demonstration site, emphasizing the significance of understanding the potential hazards posed by injection-induced seismicity in the context of carbon capture and storage (CCS). The study highlighted the U.S. Geological Survey's (USGS) monitoring efforts, which began in July 2013, consisting of 12 stations, three of which incorporated borehole sensors at depths of 150 m. The research underscored the importance of an accurate velocity model for event locations using a velocity model derived from vertical seismic profile (VSP) data collected in the CCS-1 well. The study further refined the locations of a subset of events using double-difference methods, achieving high precision in relative locations. The relocations revealed more linear patterns for seismic clusters, especially the central cluster near the injection well. Most events were identified in the granite basement, even though the injection was targeted within the lower Mt. Simon Sandstone. The paper provides valuable insights into the seismic behavior of the Decatur site and the potential implications of induced seismicity in CCS operations. For instance, the research highlighted the critical role that a meticulously derived velocity model plays in  enhancing the precision of event localization, which in turn facilitated a deeper understanding of seismic event distributions relative to the injection site. The application of double-difference methods not only refined the seismic event locations but also unveiled more defined linear patterns in seismic clusters, particularly around the central cluster near the injection well. Such patterns suggest a directional influence of the injected CO$_2$ on seismic activity, specifically that the faults are activated in certain directions, underscoring the interaction between subsurface geologic structures and injection processes. Moreover, the identification of most seismic events within the granite basement, despite targeting the lower Mt. Simon Sandstone for injection, provided pivotal insights into the subterranean migration of the injected CO2 generated pressure plume and its potential to activate seismicity in deeper geologic layers. These findings underscore the complexity of injection-induced seismicity and its implications for the safety and monitoring strategies of CCS operations.

In a related work, \cite{dando2021relocating} applied the double-difference algorithm for seismic event relocation at the Decatur site. Out of the previously located 5397 events, the research team manually picked the data to ensure its reliability. The results of the relocation showed that the seismicity of each region became less diffuse, with improved clustering in some cases. This study revealed the importance of reliable and high-quality data input for the double-difference relocation. The paper also highlighted the consistency between different data types, especially between cross-correlation derived P- and S-wave arrival time residuals. Thus, it underscored the potential of the double-difference algorithm in refining the accuracy of seismic event locations, providing invaluable insights into fault geometry, activity, and potential risks at the Decatur site.

Microseismic methods, with their rich history, have been a mainstay in the realm of geological applications \cite{bauer2019, qin2023microseismic}. Traditional techniques, such as the double-difference relocation methods, have been instrumental in refining the accuracy of seismic event locations \cite{waldhauser2000double}. The microseismic-related methods provide invaluable insights into fault geometry, activity, and potential risks~\cite{qin2022influence,langet2020, albaric2014}. Traditional traveltime-based methods such as Hypo-DD and Tomo-DD, while not computationally intensive for datasets such as the Decatur dataset, can have their own implementation complexities. The variability in inversion parameters and the subjective nature of their implementation can affect the outcomes significantly. In contrast, waveform-based methods tend to be computationally demanding, particularly when applied to large-scale datasets or complex geological settings, due to their intensive data processing requirements.This makes them less feasible for real-time, in-situ monitoring. 

In contrast, once trained, machine learning (ML) models, offer near real-time inference speed neglecting the detection step, scalability to vast datasets, and adaptability to evolving known geological conditions, this study, this adaptability is demonstrated using the Vp model. Their ability to integrate multiple data types further enhances seismic event location prediction robustness and accuracy. With minimal human oversight required post-training, ML methods, if trained correctly, present a transformative approach to seismic monitoring, promising more responsive and adaptive insights into subsurface activities.

\section{Methods: $\mathbf{\emph{DeFault}}$ Workflow}\label{sec:methods}
In recent years, machine learning has emerged as a transformative force in geophysical applications~\cite{bergen2019preface}. The vast and complex datasets inherent to geophysics, from seismic traces to subsurface imaging, have found a natural ally in ML algorithms. These algorithms excel at identifying patterns, anomalies, and relationships within large-scale, multi-dimensional data. Once trained, the neural network models can rapidly process and interpret data, offering insights that might be overlooked by conventional methods. From enhancing seismic inversion techniques to automating fault detection and improving subsurface characterization, ML is paving the way for more efficient, accurate, and comprehensive geophysical analyses. As the geoscience community continues to embrace the potential of ML, it stands poised to unlock deeper understanding and more innovative solutions in the exploration and monitoring of Earth's subsurface.

Thus, in this study, we propose to use a machine learning workflow that can delineate faults and fractures induced by CO$_2$ injections at the Decatur site as a case study. The proposed workflow takes the detected seismic monitoring data profiles without further signal processing as input images, converts the recorded passive seismic events to their source spatial location, and finally utilizes the source locations to delineate the fault structure. We focus on the P wave arrivals recorded by the 31 vertically distributed receivers being placed in a borehole well GM-1 above the reservoir layer and 2 receivers at the injection ports in the CCS-1 well. 

The workflow can be divided into four main steps:
\begin{enumerate}
    \item \textbf{Signal processing and enhancement:} We first apply a series of filters to the raw field seismic recorded waveforms to remove noise and enhance effective signals, specifically, to improve the signal-to-noise ratio (SNR).
    \item \textbf{Deep-learning-based event location:} A large synthetic training set is generated, and combined with the processed field data to be fed into an MLReal domain adaptation encoder-decoder neural network training stage.
    \item \textbf{K-means clustering:} The processed field data pass through the MLReal steps and is fed into the trained encoder-decoder network to predict microseismic event locations. These predictions within a certain time period are then clustered with a K-means algorithm, which automatically separates the events spatially. A least square distance match is then applied to find the corresponding fault plane in each cluster.
    \item \textbf{Dropout uncertainty analysis:} Dropout uncertainty analysis involves modifying a neural network by inserting dropout layers, which randomly ignore a subset of neurons during training. This technique often prevents overfitting~\cite{srivastava2014dropout,6639346}, and thus, promotes a model that can generalize better to unseen data. By applying dropout during testing (not just training), we can measure the variability in the network's predictions, offering a practical approach to gauge the model's confidence in its predictions. Essentially, this method allows us to assess the robustness of the neural network's outputs, which is crucial for understanding the potential range of outcomes and making informed decisions based on the model's predictions. We modify and retrain the network architecture by adding dropout layers with dropout rate $p=0.2$ after each convolutional layer except the last one, and use it to perform uncertainty analysis. 
\end{enumerate}
Details of each step will be discussed in the following subsections. 

\begin{figure*}[!ht]
\centering
\begin{subfigure}[b]{0.31\columnwidth}
    \centering
    \includegraphics[trim={3.5cm 0cm 4cm 1.3cm},clip,width=1\columnwidth]{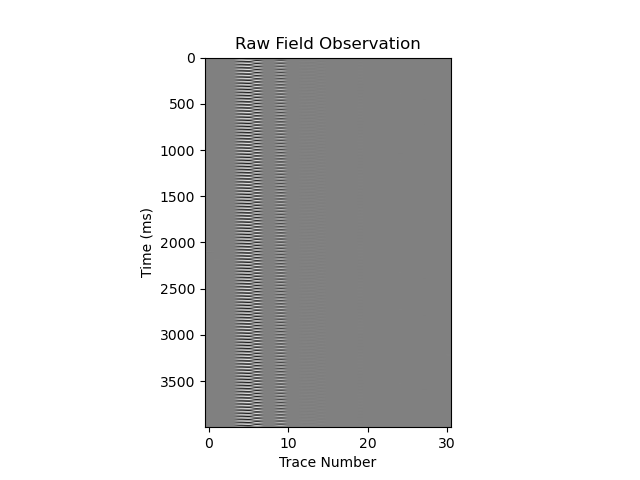}
    \caption{Raw field data}
    \label{fig:raw_field}
\end{subfigure}
\begin{subfigure}[b]{0.31\columnwidth}
    \centering
    \includegraphics[trim={3.5cm 0cm 4cm 1.3cm},clip,width=1\columnwidth]{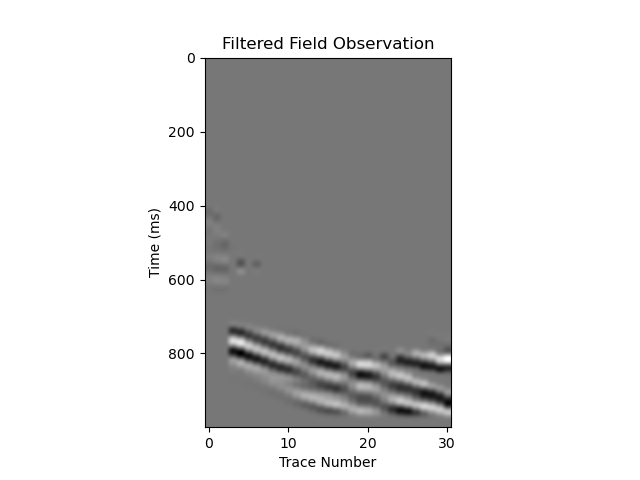}
    \caption{Filtered field data}
    \label{fig:filter_field}
\end{subfigure}
\begin{subfigure}[b]{0.31\columnwidth}
    \centering
    \includegraphics[trim={3.5cm 0cm 4cm 1.3cm},clip,width=1\columnwidth]{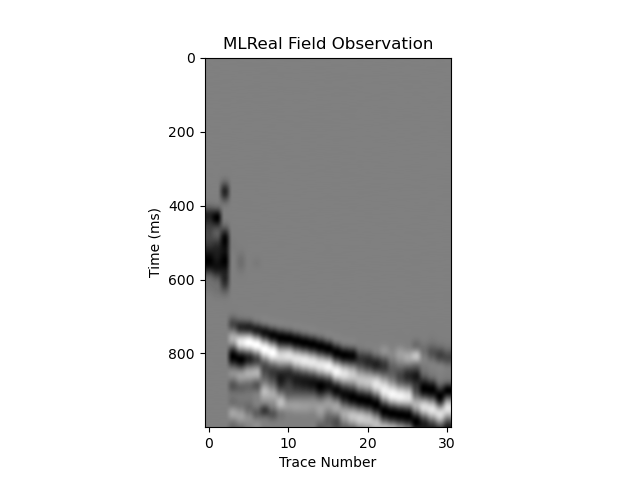}
    \caption{MLReal field data}
    \label{fig:mlr_field}
\end{subfigure}
\begin{subfigure}[b]{0.31\columnwidth}
    \centering
    \includegraphics[trim={3.5cm 0cm 4cm 1.3cm},clip,width=1\columnwidth]{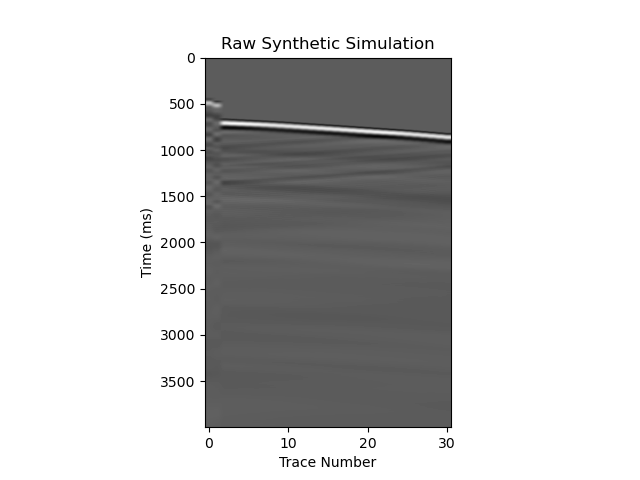}
    \caption{Simulated synthetic data}
    \label{fig:raw_synthetic}
\end{subfigure}
\begin{subfigure}[b]{0.31\columnwidth}
    \centering
    \includegraphics[trim={3.5cm 0cm 4cm 1.3cm},clip,width=1\columnwidth]{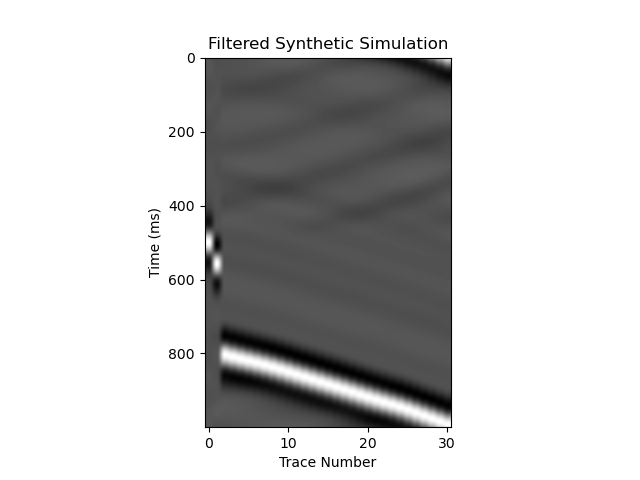}
    \caption{Filtered synthetic data}
    \label{fig:filter_synthetic}
\end{subfigure}
\begin{subfigure}[b]{0.31\columnwidth}
    \centering
    \includegraphics[trim={3.5cm 0cm 4cm 1.3cm},clip,width=1\columnwidth]{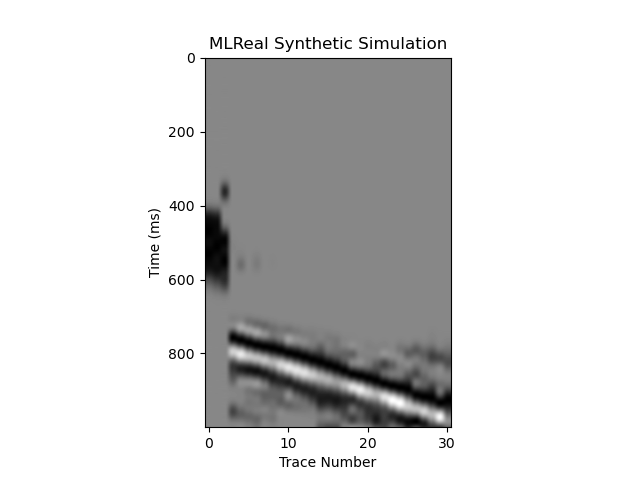}
    \caption{MLReal synthetic data}
    \label{fig:mlr_synthetic}
\end{subfigure}
\caption{Field and synthetic data comparison before and after processing. Field data (a) from raw observation, (b) after filtering, (c) after MLReal steps for the inference stage, and synthetic data (d) simulated, (e) temporally cropped, (f) after MLReal steps for the training stage. Vertical axis: time (ms); horizontal axis: trace number.}
\label{fig:process}
\end{figure*}

\subsection{Signal processing and data enhancement}\label{sec:processing}

For setting the foundation for our study, we use the detected seismic event data from IBDP. A sample of a raw seismic record from this source, presented in Figure~\ref{fig:raw_field}, is deteriorated by an obvious low signal-to-noise ratio (SNR). This compromised state of raw data can prevent efforts of accurately detecting effective seismic events. To overcome this issue and ensure optimal inputs for our neural network training, we engage in a robust preprocessing scheme designed for the Decatur IBDP seismic monitoring data, thoroughly preparing it for subsequent phases of neural network training and inference.

\begin{figure*}[!ht]
\centering
\begin{subfigure}[b]{0.46\columnwidth}
    \centering
    \includegraphics[trim={0cm 6cm 0cm 6cm},clip,width=1\columnwidth]{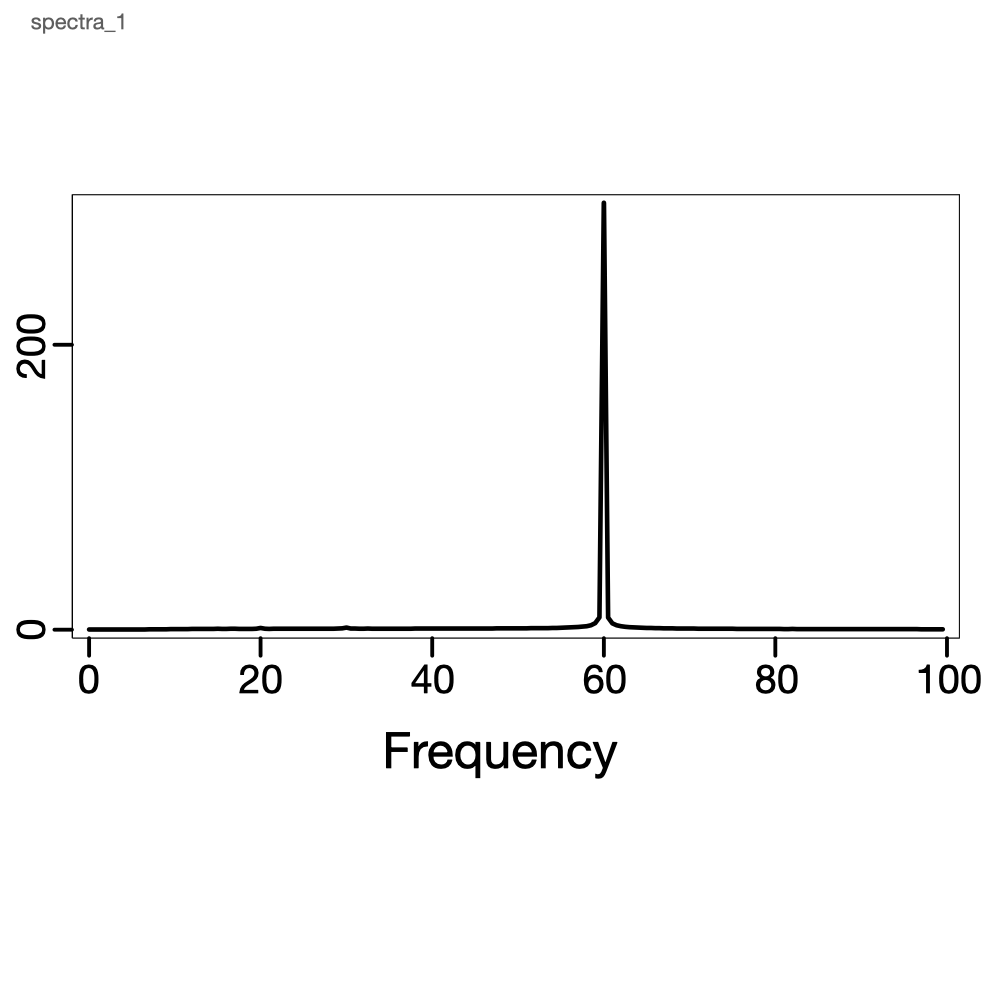}
    \caption{Raw data frequency spectrum}
    \label{fig:spectrum_1}
\end{subfigure}
\begin{subfigure}[b]{0.46\columnwidth}
    \centering
    \includegraphics[trim={0cm 6cm 0cm 6cm},clip,width=1\columnwidth]{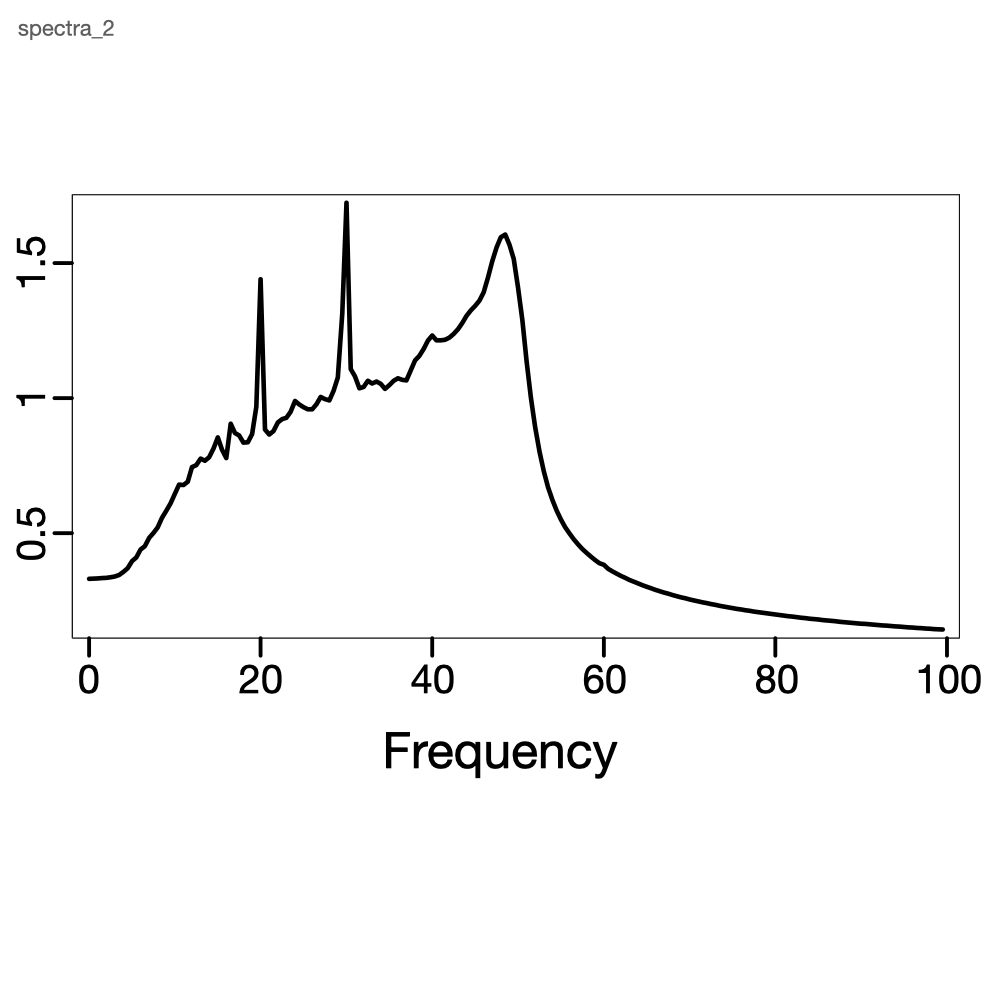}
    \caption{Post-lowpass frequency spectrum}
    \label{fig:spectrum_2}
\end{subfigure}
\caption{Noise component frequency spectra to be filtered out. (a) Frequency spectrum of the raw data revealing a marked 60Hz energy intrusion. (b) Frequency spectrum of the lowpass filtered data revealing the noise of 20 and 30Hz.}
\label{fig:spectra}
\end{figure*}

\textbf{1. Lowpass Filter:} A preliminary study of the raw data's stacked frequency spectrum, illustrated in Figure~\ref{fig:spectrum_1}, unveils a dominant noise at 60Hz, which is caused by EM signals from the power grid being picked up by the geophones. To address this noise, we employ an initial lowpass filter, selectively preserving frequencies below 50 Hz. This action is pivotal in mitigating the influence of high-frequency noise, such as the 60Hz noise, guiding the model's focus toward the seismic event's essential frequency range.

\textbf{2. Notch bandpass filters:} Following the initial filtering process, the data encounters a succession of notch bandpass filters, strategically designed to eliminate interfering frequencies around 20 Hz and 30 Hz, as illuminated in Figure~\ref{fig:spectrum_2}. This precise filtration purges artifacts linked to geophone instrumental noise, protecting the SNR of data for the network's subsequent feature extraction phase. The notch filters are specially designed for the IBDP data, and can be ignored or re-designed accordingly to other site data.

\textbf{3. Trace-wise amplitude normalization:} To ensure a consistent amplitude spectrum across seismic traces and geophones, we apply a trace-wise normalization within the fixed 4-second detection windows provided by the Decatur data preprocessing steps, where the detected event is centered. Each seismic trace is scaled such that peak absolute values are normalized between $[-1, 1]$. This adjustment allows the encoder-decoder network to train unaffected by arbitrary trace amplitudes. While our algorithm prioritizes event relocation and does not utilize amplitude information for this purpose, it is important to note that this normalization step may obscure source mechanism information.

\textbf{4. F-k domain dipping filter:} The F-k domain dipping filter operates in the frequency-wavenumber (F-k) spectrum, a method used in signal processing to analyze and manipulate signals based on their frequency and wavenumber components, where frequency indicates the rate at which a wave oscillates and wavenumber is a measure of spatial frequency indicating the number of wave cycles per unit distance. This technique filters out energy that does not conform to expected geological formations by focusing on specific frequency and wavenumber ranges. Parameters for this filter include a frequency cutoff of 50 Hz and a wavenumber limit tailored to enhance the SNR by attenuating noise associated with frequencies above 60 Hz, which are typical of electrical interference and other ambient noise. We incorporate an F-k domain dipping filter, specifically designed to suppress incongruous dipping noise phenomena within the seismic records. Operating in the frequency-wavenumber (F-k) spectrum, this filter discriminately removes energy inconsistent with predicted geological formations, thus enhancing the overall SNR.

\textbf{5. Time domain anomalous noise purge:} To mitigate the impact of random noise manifestations, such as sudden spikes or transient energy boosts, we implement a time domain filter. This filter operates within a 300 ms window for each trace, adjusted according to the vertical positioning of the receivers in the borehole, with the detected event centered in this window. The time domain filter is designed to suppress noise outliers that could impair the learning capacity of our network, thereby preserving data quality and enhancing model reliability.

\textbf{6. F-k domain envelop filtering:} Concluding our signal enhancing process, we use an F-k domain filter constructed on the summation of the F-k domain spectrum envelope of the synthetic training samples. By overlaying this filter on the field data, we ensure suppression of residual noise, refining the data for the neural network training phase.

Emphasizing repeatability, each of these signal enhancement procedures is carefully standardized, setting the stage for a seamless, automated workflow that necessitates minimal manual interactions. Through this preprocessing, we endow our neural network with streamlined, harmonized inputs. Such a holistic approach significantly elevates the network's feature extraction power, which will help improve the final performance of the network on the field seismic data.

\subsection{Deep-learning-based event locating}\label{sec:relocation}
\begin{figure*}[!ht]
\centering
\includegraphics[width=0.8\columnwidth]{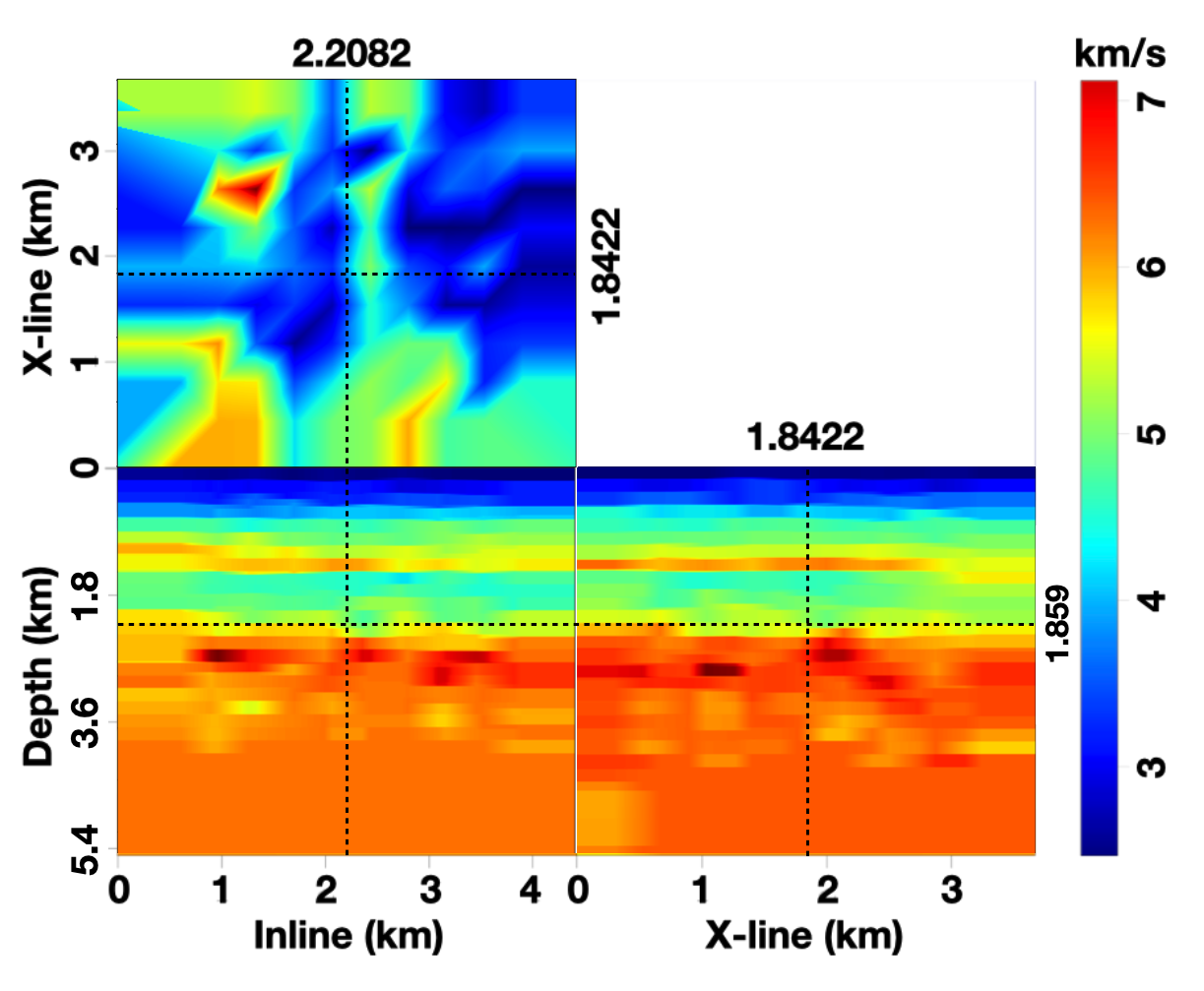}
  \caption{Interval velocity transformed from the stacking velocity model of the 2011 active seismic survey in the region. It is used for generating the synthetic training data. }
\label{fig:internal}
\end{figure*}

In our methodology, we leverage ML to interpret seismic data. At its core, our model analyzes patterns within the seismic data, which is the seismic event locations, to identify fault lines. By training our model on both synthetic and real-world data, we ensure that the algorithm learns the complex signatures indicative of geological formations. This training will enable the model to predict fault locations with a degree of accuracy previously unattainable through conventional analysis alone, as shown in the following sections. In this section, we will breakdown the steps on how a deep-learning neural network can extract useful features from the processed seismic data and relocate the events to their original source locations.

Training neural networks typically demands extensive training sets, which should ideally align with the distribution of the test sets. We, therefore, use a 3D interval velocity model, derived from stacking velocity measurements done in 2011, and randomly sample 15,000 potential microseismic sources from the area we expect such sources reside, to generate 15,000 random synthetic seismic data samples, with the event locations as labels. These samples use the IBDP monitoring system geometry, and a Ricker wavelet with $30$Hz peak frequency as the source function to align with the field data average frequency spectrum. The velocity model is displayed in Figure~\ref{fig:internal}, with a sample of simulated seismic data portrayed in Figure~\ref{fig:raw_synthetic} using the Madagascar open-source project code \cite{fomel2013madagascar}. Since the focus is on using direct microseismic arrivals, and other arrivals tend to be separable with the P arrivals, we use the transmission component of the wavefield, which can be modeled reasonably accurately with velocities extracted from stacking velocities.

A significant challenge in leveraging neural networks trained on seismic waveform data is their real-world applicability. Due to the necessity for precise labeling, synthetic data, where labels are immediately accessible, often becomes the primary training resource. However, such synthetic data might not capture the nuances of field experiments, leading to the underperformance of trained neural network models during inference. Real-world data encompasses features not present in synthetic datasets, such as precise waveform source signatures, field noise, and the subsurface reflectivity.

\subsubsection{Bridging the synthetic-real data discrepancy for field data applications}\label{sec:mlr}
Our study introduces a new approach, MLReal, for seismic event adaptation~\cite{alkhalifah2022mlreal}. This method proficiently overcomes the limitations of traditional noise-sensitive techniques and the substantial data distribution prerequisites of conventional machine learning (ML) models.

MLReal emerges as a compelling domain adaptation technique, designed to bridge the gap between the distributions of the synthetic training data used in machine learning and the field data. By subjecting synthetic data to a sequence of linear operations, it mirrors the intrinsic source functions and noise differences in field data, consequently improving the training effectiveness and performance of Convolutional Neural Networks (CNN).

For this methodology, we make an assumption: the application data is devoid of labels, rendering transfer learning, a recognized domain adaptation approach, infeasible. In this study, we define the \emph{source} set as the synthetic labeled dataset, with the paired simulated seismic waveforms and the corresponding ground truth labels; the \emph{target} set as the real field unlabeled dataset, with the noisy field observed seismic waveforms, whose labels are our objective in the inference. In our neural network model, the input data to the network is the synthetic and field waveform data in the training and inference stages, respectively. The output is the predicted seismic event locations, correspondingly. Such a domain adaptation scenario is typically navigated using unsupervised ML strategies. Within this context, an essential premise is that both target and source labels share identical distributions: \(P({\bf y_s})=P({\bf y_t})\), where $y$ refers to output labels, $P(\cdot)$ refers to the probability distribution, and the subscripts $_s$ and $_t$ refer to the source and target, respectively. Such a parallelism is indispensable when deploying synthetically trained neural network (NN) models on actual data. It presupposes that tasks executed on field data are echoed by the synthetic training data, denoted by \(P_s({\bf y_s}|{\bf x_s})=P_t({\bf y_t}|{\bf x_t})\), where $x$ is input data. Our method ensures that the synthetic data we use for training closely imitates the real-world physical behavior and conditions, like the geological features of the Earth. Essentially, we're saying that our model's assumptions are closely aligned with what happens in the real world. This means that when our neural network processes the data, the results it generates take into account the real-world constraints and conditions, making the predictions more reliable and accurate.

A significant challenge arises when the input distributions for source and target data are not aligned or similar to each other, particularly in scenarios where \(P_s({\bf x_s}) \neq P_t({\bf x_t})\). This deviation between the training and application dataset distributions is termed `covariate shift' in the realm of machine learning. Numerous tools, such as the Kullback-Leibler (KL) divergence metric, can quantify such a shift. 
Our goal is to use a transformation that contracts the difference in distribution between training/synthetic and application/field datasets \cite{alkhalifah2022mlreal}.

For microseismic monitoring, we assume that the recorded data are sorted in temporal sections corresponding to detected microseismic events, with each section, indexed by $j$, containing recorded traces, $d^{ij}(t)$, over time, $t$, and indexed by the trace number, $i$, corresponding to the location of the sensors.
Thus, for a synthetically generated training set, we follow a similar configuration. That is, we simulate a section of seismic data, indexed by $d_s^i(t)$, by assuming that each of the microseismic events are randomly triggered spatially. We drop the section index, $j$, as we focus on the transformation of one section (recorded event). The MLReal transformation involves the following operations to obtain a transformed synthetic test trace: 

\begin{equation}
\hat{d^i_s}(t) = d_s^i(t) \otimes d_s^k(t) \ast d^{ij}(t) \otimes d^{ij}(t).
\label{eq:syn2fd}
\end{equation}

The index \( k \) denotes a constant reference trace from synthetic data within the same section. The symbol \(\ast\) means the convolution operation, while the symbol \(\otimes\) means cross-correlation. Note that the correlation operation is actually an autocorrelation of the field data, which yields a zero-phase output that mainly includes the source signature and noise information from the field data that are transferred to the synthetic training data. In the training, the section, $j$, is drawn randomly from the field dataset per epoch of training data, in which we allow for the transfer of a distribution of field data samples features to every synthetic trace.

We reciprocate the operations for the inference stage. However, since we do not have the luxury of incorporating random synthetic data sections into the field data traces (since inference is a feed forward operation, no epochs), we actually incorporate the mean of the autocorrelation of the field data. Thus, the resulting inference field data trace is given by
\begin{equation}
\hat{d^i_t}(t) = d^i_t(t) \otimes d^k_t(t) \ast \frac{1}{N_s} \sum_j d_s^{ij}(t) \otimes d_s^{ij}(t).
\label{eq:fd2syn}
\end{equation}
On the right-hand-side (RHS) of equations~\ref{eq:syn2fd}~and~\ref{eq:fd2syn}, the operations from left to right suggest that 1) the sample trace from a section is first cross-correlated with a fixed reference trace from that section to pull the event waveform to the zero-time-lag region because the onset of raw recorded seismic events are unknown; 2) the correlated trace is then convolved with the other dataset (inference or training for equations~\ref{eq:syn2fd}~and~\ref{eq:fd2syn}, respectively) to incorporate its features; 3) The autocorrelation admits a zero-phase signal carrying mainly the noise, source signature, and reflectivity information (specifically for the real data) information of the autocorrelated data. 

This transformation essentially bridges the gap between source and target data distributions. The resultant training and inference data sets facilitate the neural network's more efficient operation on real data, enabling the model to adeptly recognize patterns in both. The enriched synthetic data set now more closely mirrors real-world conditions, positioning it as an ideal training data set for the neural network model.

\subsubsection{Neural network architecture: an encoder-decoder approach}\label{sec:network}

\begin{figure*}[!ht]
\centering
\includegraphics[width=1\columnwidth]{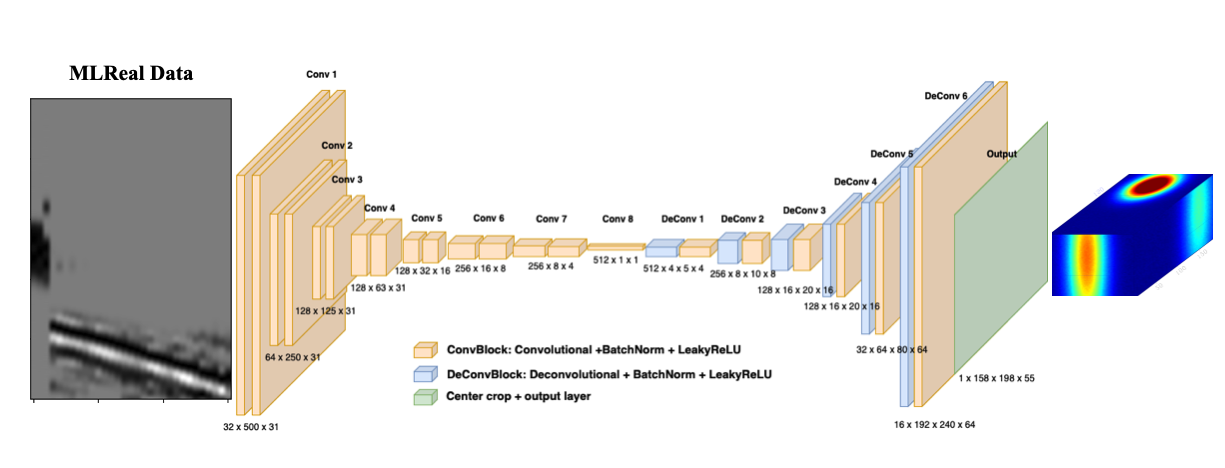}
  \caption{The architecture of the neural network employed for passive seismic event relocation. It ingests processed seismic data and emits a heat-map indicative of event location distribution.}
\label{fig:network_architecture_heatmap}
\end{figure*}

Traditional approaches to relocating passive seismic events might utilize architectures reminiscent of VGG-16, a famous classic CNN architecture for computer image processing~\cite{simonyan2015deep}, whereby seismic events serve as inputs, culminating in event location coordinates as outputs~\cite{alkhalifah2022mlreal,wang2021data,wang2021direct}. However, in computer vision, heat-maps have established themselves as invaluable assets, especially in tasks like semantic segmentation~\cite{long2015fully}, human pose estimation~\cite{newell2016stacked,cao2017realtime}, and saliency prediction~\cite{zhang2016exploiting}. Beyond visualization, heat-maps accentuate specific image regions, lending models enhanced interpretability by aligning with distinct classifications~\cite{zhou2016learning,springenberg2014striving,selvaraju2017grad}.

For enriched performance on the transformed synthetic seismic dataset, we propose taking advantage of heat-map techniques with an encoder-decoder based training approach. This synergistic strategy not only capitalizes on the MLReal transformations, but also leverages the encoder-decoder's adeptness in distilling salient features, culminating in output heat-maps ripe for juxtaposition against ground truth labels.

The encoder-decoder architecture consists of a two-phase approach: encoding the input data into a latent representation vector and then decoding it. Applied herein, it adeptly handles synthetic seismic data processed via MLReal. The encoding phase extracts the input into a concise representation, whereas decoding unfolds this representation. A visual breakdown of the architecture used in this study is shown in Figure~\ref{fig:network_architecture_heatmap}.

Consequently, the network outputs a 3D heat-map with dimensions $[X,Y,Z]=[158,198,55]$. This output can be compared under the $\ell 2$ norm loss function with heat-maps, derived from labels via 3D Gaussian functions orbiting the true source location coordinates. The training phase engages the minimization of such a loss function with an Adam optimization strategy~\cite{kingma2017adam}.

To accurately derive source coordinates from the predicted heat-maps, our process involves two main steps. First, we upscale the low-resolution 3D heat-maps to higher resolution using spline interpolation. This step eliminates the grid effects often observed with passive seismic source detections. Next, we identify the highest (peak) value and the second highest (sub-peak) value within these refined heat-maps. The final source coordinates $(x, y, z)$ are calculated by adjusting the position from the sub-peak towards the peak, as quantified in the equation:

\[
(x, y, z) = 0.25 \times (\text{peak position}) + 0.75 \times (\text{sub-peak position})
\]

This method ensures precise localization of seismic events by leveraging the spatial information in the heat-map's intensity peaks.

In summation, the training strategy takes advantage of the encoder-decoder model and a heat-map-centric methodology, enhancing the network's performance in extracting significant features from the transformed synthetic seismic data. This combination of training strategies, when coupled with MLReal transformations, paves the way for a state-of-the-art model exhibiting commendable adaptability to real-world seismic data, affirming its generability across multiple tasks.

\subsection{K-means clustering for fault delineation}\label{sec:clustering}

Fault and fracture delineation using passive seismic data demands intricate geospatial analysis. After identifying the spatial locations of passive events, employing an event clustering algorithm combined with fault delineation becomes essential. This procedure is designed to highlight potential faults and fractures throughout the targeted area. Central to this approach is the clustering of seismic events by their geospatial characteristics — latitude, longitude, and depth — alongside the recording timestamp. For this purpose, the K-means algorithm stands out as the method of choice.

The first step, termed \textit{event temporal grouping}, involves classifying seismic events according to their occurrence time-frame, which can vary from broader periods like years to more detailed spans such as months. Subsequently, in the \textit{K-means clustering execution} phase, the K-means algorithm is applied to each temporal segment, focusing on the geographical positions of events. For segments registering more than five events, the number of clusters is determined based on the event count, with at least five clusters set as a baseline. The K-means algorithm operates iteratively, partitioning \( n \) data points into \( k \) distinct, non-overlapping clusters, in an unsupervised learning manner. It achieves this by initially positioning \( k \) centroids randomly. Each data point is then associated with the nearest centroid. Subsequently, centroids are recalibrated to represent the mean of their associated data points. This iterative process continues until centroid movement between iterations becomes negligible. Through this method, seismic events are clustered by their geographical coordinates and depth, offering insights into event patterns over set intervals. The last phase, named \textit{fault plane configuration}, closely examines each cluster. Outliers situated more than 0.5km from a cluster's epicenter are excluded, based on the assumption that they likely belong to a different fault line than the cluster's predominant one. After this exclusion, a plane, and its strike line, is drawn through the cluster employing a least squares method.

This algorithmic process results in a series of 3D visualizations, with each representation correlating to a distinct time segment, illustrating the spatial clustering of seismic events within. Comprehensive results, covering long-term (1440 days) clustering and delineations, are presented in the following section.

\subsection{Uncertainty quantification and dropout analysis}\label{sec:uq}

Uncertainty quantification (UQ) remains a critical facet of deep learning models, especially when the uncertainty is high, as in the geospatial analysis with the $\mathbf{\emph{DeFault}}$ algorithm. To comprehensively analyze the inherent uncertainty and ascertain the reliability of our network's predictions, we perform an in-depth analysis based on the Decatur dataset.

Inspired by the work of \cite{kendall2017uncertainties}, in which the authors study the benefits of modeling epistemic vs. aleatoric uncertainty in Bayesian deep learning models for vision tasks, our methodology involved a nuanced adjustment to the original network architecture. Specifically, we introduce dropout layers to stochastically deactivate a subset of neurons during training, thereby injecting randomness into the model. Each of these layers had a dropout rate of \( p=0.2 \) and was positioned after every convolutional layer, with exception to the last output layer.

After training the modified network, we subjected the $\mathbf{\emph{DeFault}}$ algorithm to the processed Decatur dataset to examine its performance. Notably, this evaluation wasn't a one-off experiment. Instead, we repeated the inference step 50 times to capture the variability in the model's output, caused by the incorporated dropout layers. By computing the mean of these 50 evaluations, we established a ``reference location'' for any given seismic event. More profoundly, the standard deviation across these predictions furnished a quantitative measure of the uncertainty tied to each prediction, thus offering a clearer understanding of the trustworthiness of our model's predictions.

\section{Results and Analysis}\label{sec:results}

We evaluated the $\mathbf{\emph{DeFault}}$ method using the Decatur CO$_2$ storage project, a pivotal initiative under the Department of Energy's carbon capture and storage program. This project aimed to demonstrate the potential to mitigate climate change by securely storing commercial volumes of CO2 in the subsurface. Figure~\ref{fig:decatur1} provides a map of the Decatur site in Illinois. The project successfully demonstrated the potential of large-scale CO$_2$ storage in deep saline reservoirs and contributed valuable insights into the storage mechanism.
\begin{figure*}[!ht]
\centering
\includegraphics[trim={3cm 4cm 5cm 3cm},clip,width=0.5\columnwidth]{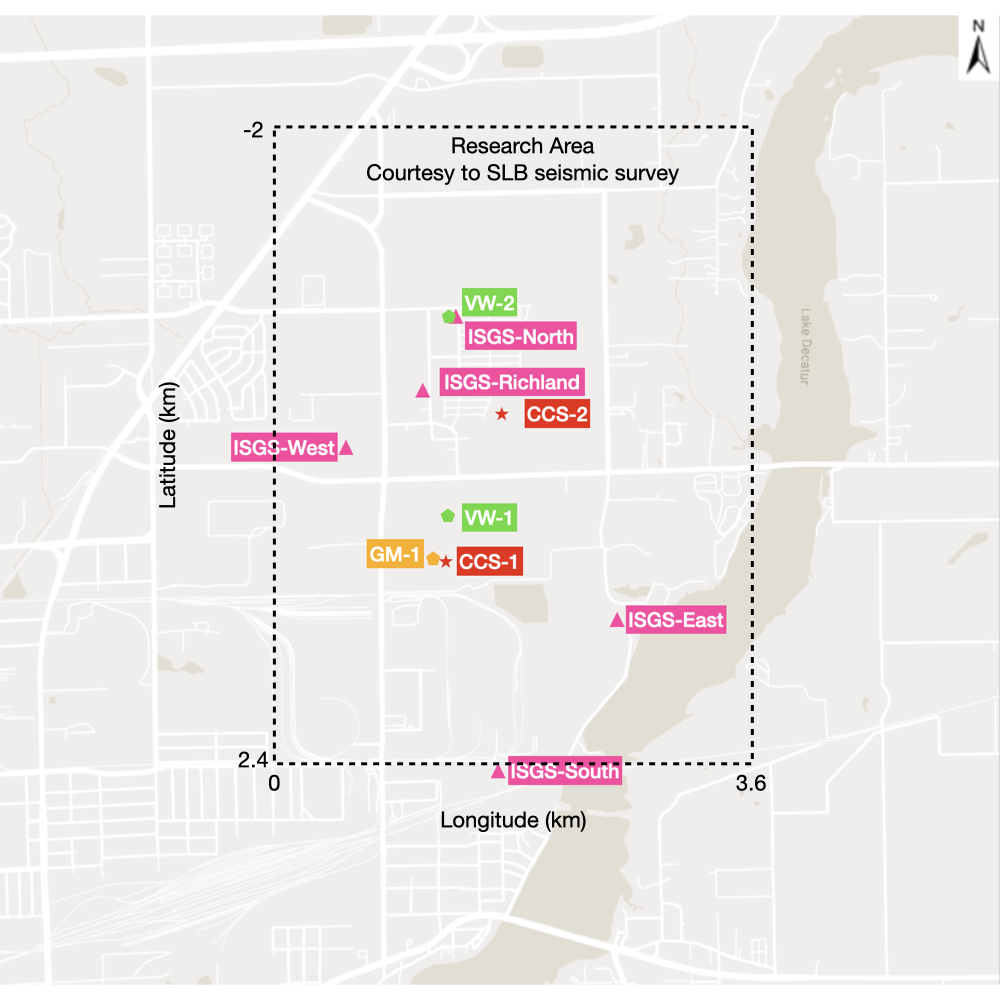}
  \caption{The locations of the monitoring stations and injection wells at the Decatur CO$_2$ site, including (magenta dots) ISGS surface stations, (orange dots) GM-1 downhole geophones, (green dots) VW downhole geophones, and (red dots) CCS-1 $\&$ -2 injection wells and geophones.}
\label{fig:decatur_geometry}
\end{figure*}
Our analysis leveraged data from the Illinois Basin-Decatur Project (IBDP), which are publicly accessible on the NETL's EDX platform. Currently, the project is monitored using borehole sensors complemented by five ISGS surface stations and several other USGS stations. Our primary focus remains on well data due to the strategic sensor arrangement and its ability to provide the relatively high-quality moveout features of the passive events.

The dataset comprises the recorded traces from 31 sensors: 2 from CO$_2$ injection CCS-1 well and 29 from the GM-1 well. Figure~\ref{fig:decatur_geometry} illustrates the positioning of the injection and observation wells, represented by orange and red stars, respectively. As we operate in a three-dimensional framework with only a $P$-wave stacking velocity model on hand, and considering the resource-intensive of performing large-scale 3D synthetic simulations, we choose to simplify the wave phenomena by only considering the acoustic effect and ignoring the elastic one. Using a finite difference acoustic wave equation solver with a constant density assumption, we synthetically generated a comprehensive training set, which is discussed in section~\ref{sec:relocation}. These locations are constrained to a depth range of $[1.7, 2.4]$ km to match the CO$_2$ reservoir depth and the basement beneath. The depth range is  designed to reasonably simulate the expected reservoir layers. The synthetic source locations serve as training labels when converted to heat-map format, as detailed in Section~\ref{sec:network}. To synchronize with the acoustic synthetic training data, we extracted only the Z-component recordings from the multi-component geophones of the CCS-1 and GM-1 wells, because the Z-component tends to capture upcoming P-waves, and those P-waves can be approximated by the acoustic assumption.

\begin{figure*}[!ht]
\centering
\begin{subfigure}[b]{0.31\columnwidth}
    \centering
    \includegraphics[trim={6cm 6cm 7cm 9cm},clip,width=1\columnwidth]{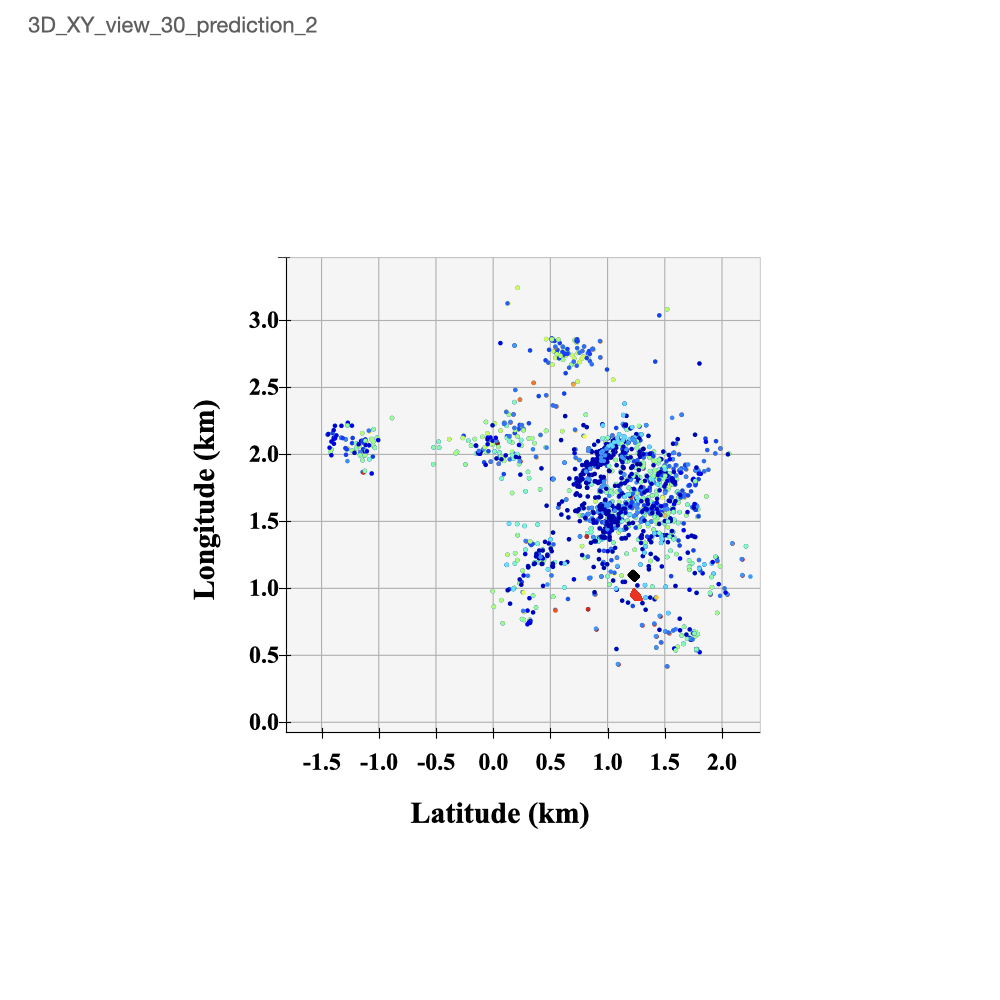}
    \caption{Longitude-latitude view. }
    \label{fig:3D_XY_view_prediction}
\end{subfigure}
\begin{subfigure}[b]{0.31\columnwidth}
    \centering
    \includegraphics[trim={7cm 8.5cm 7cm 6cm},clip,width=1\columnwidth]{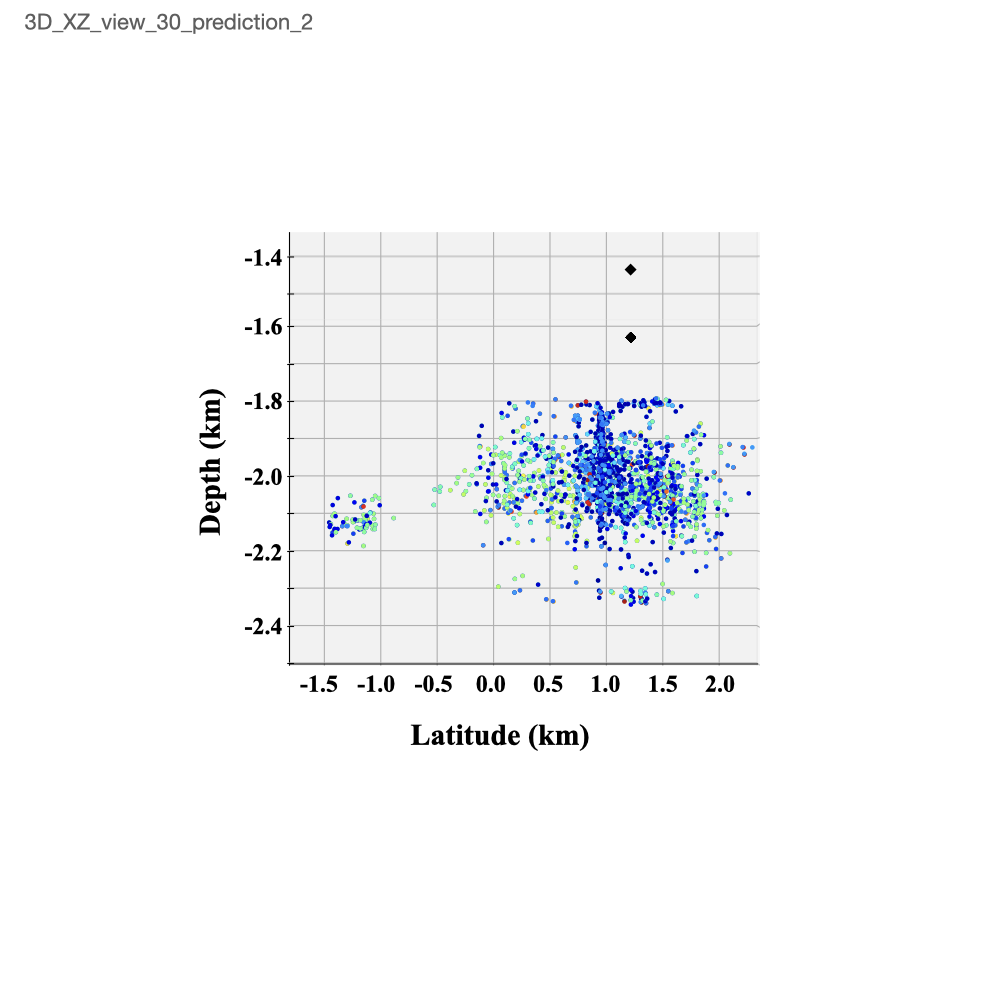}
    \caption{Latitude-depth view. }
    \label{fig:3D_XZ_view_prediction}
\end{subfigure}
\begin{subfigure}[b]{0.31\columnwidth}
    \centering
    \includegraphics[trim={7cm 8.5cm 7cm 6cm},clip,width=1\columnwidth]{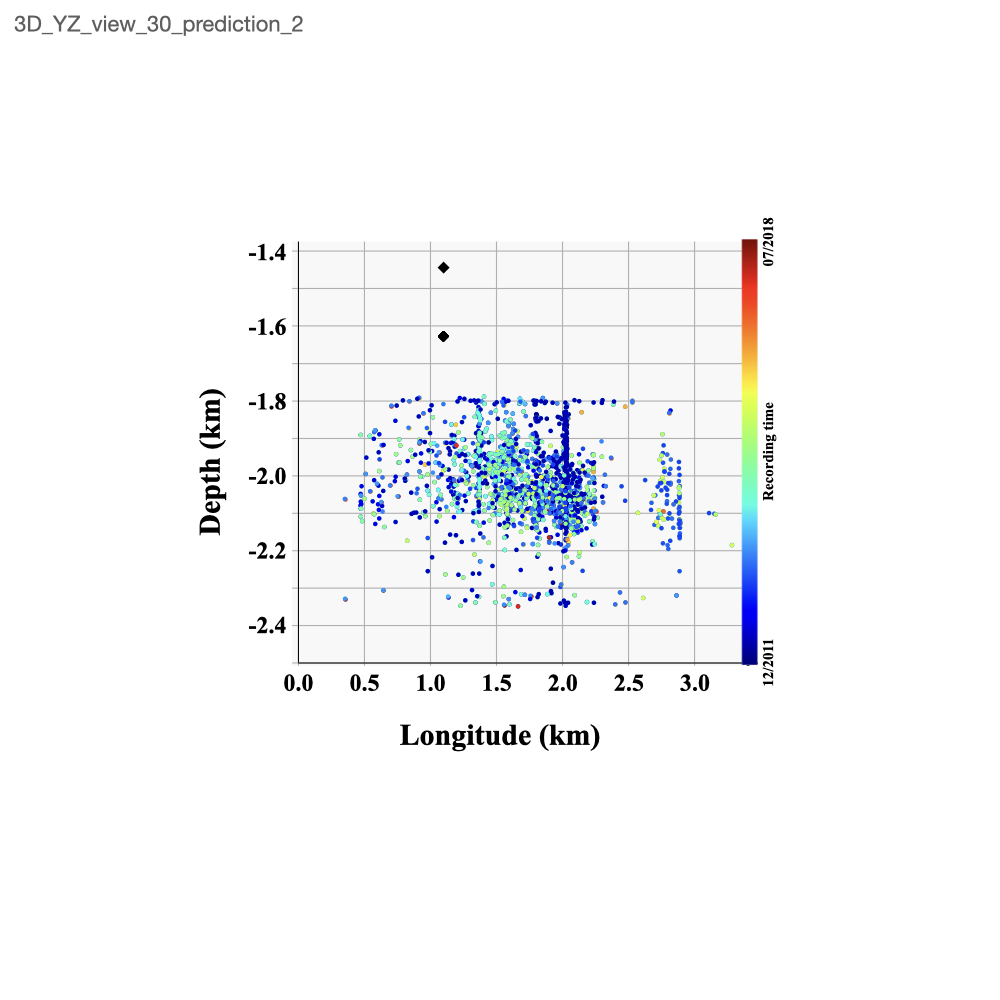}
    \caption{Longitude-depth view. }
    \label{fig:3D_YZ_view_prediction}
\end{subfigure}
\caption{$\mathbf{\emph{DeFault}}$ predicted passive seismic event locations, interpolated from the output heat-maps. (a) Longitude-latitude, (b) latitude-depth, and (c) longitude-depth views. The locations are interpolated by the maximum and second maximum values of the heat-maps. The red triangles are the GM-1 borehole receivers. The black diamonds are the CCS-1 injection ports.}
\label{fig:prediction}
\end{figure*}
Subsequently, the field data underwent the signal processing steps outlined in Section~\ref{sec:processing}. Both the processed field data and synthetic data were transformed using the MLReal method as described in Sections~\ref{sec:mlr}. The selection of synthetic-field data pairs is based on a random strategy. The 15,000 MLReal synthetic samples were randomly divided into the training (12,000 samples) and validation (3,000 samples) sets. In machine learning, separating data into training and validation sets is crucial for developing a model that performs well not just on the data it has seen in the training, but also on new, unseen data. The training set is used to teach the model how to recognize patterns and make predictions. The validation set, however, is needed to test the model's accuracy on data it hasn't been trained on, suggesting it can generalize within the training distribution, and not overfit the data. In other words, this process helps in fine-tuning the model's parameters for optimal performance while avoiding overfitting, where the model learns the training data too well and performs poorly on new data. After training our neural network using this dataset, we tested it against the processed field data. The predicted outcomes are presented in Figure~\ref{fig:prediction}, where event locations are depicted by colored dots, transitioning from blue to red to signify time progression. GM-1 sensors and CO$_2$ injection sites are denoted by red triangles and black diamonds, respectively. The predictions highlight the clustered event locations with a predominant cluster north of the injection well and a few smaller clusters in the far north-east. Events at the top and bottom boundaries indicate potential training gaps. These events may be induced at a depth that is out of the training sample depth range, and thus, the trained network cannot predict them. This implies that there are unexpected events that is located beyond the region of interest, according to our algorithm. It is reasonable as not all the monitored events are induced by injection at the reservoir layer. We can simply ignore those out-of-region events by considering them not to be induced passive events. Another possible reason is that the events are too noisy, where the effective signal energy is not recognized by the network. These events should not originate from the phase conversion involved in the MLReal steps, as we are using pre-detected events that are supposed to have effective seismic waveforms centered around the time axis.
\begin{figure*}[!ht]
\centering
\begin{subfigure}[b]{0.31\columnwidth}
    \centering
    \includegraphics[trim={6cm 6cm 7cm 9cm},clip,width=1\columnwidth]{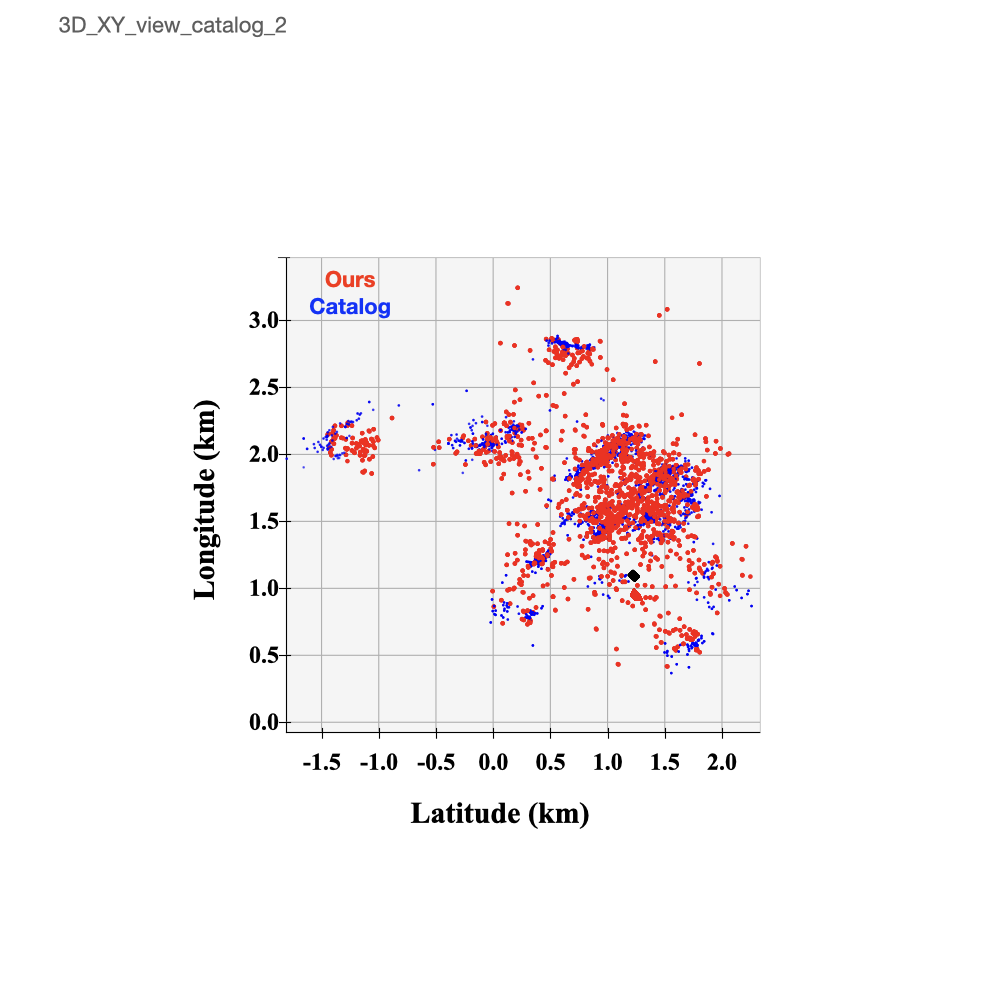}
    \caption{Longitude-latitude view. }
    \label{fig:3D_XY_view_catalog}
\end{subfigure}
\begin{subfigure}[b]{0.31\columnwidth}
    \centering
    \includegraphics[trim={7cm 8.5cm 7cm 6cm},clip,width=1\columnwidth]{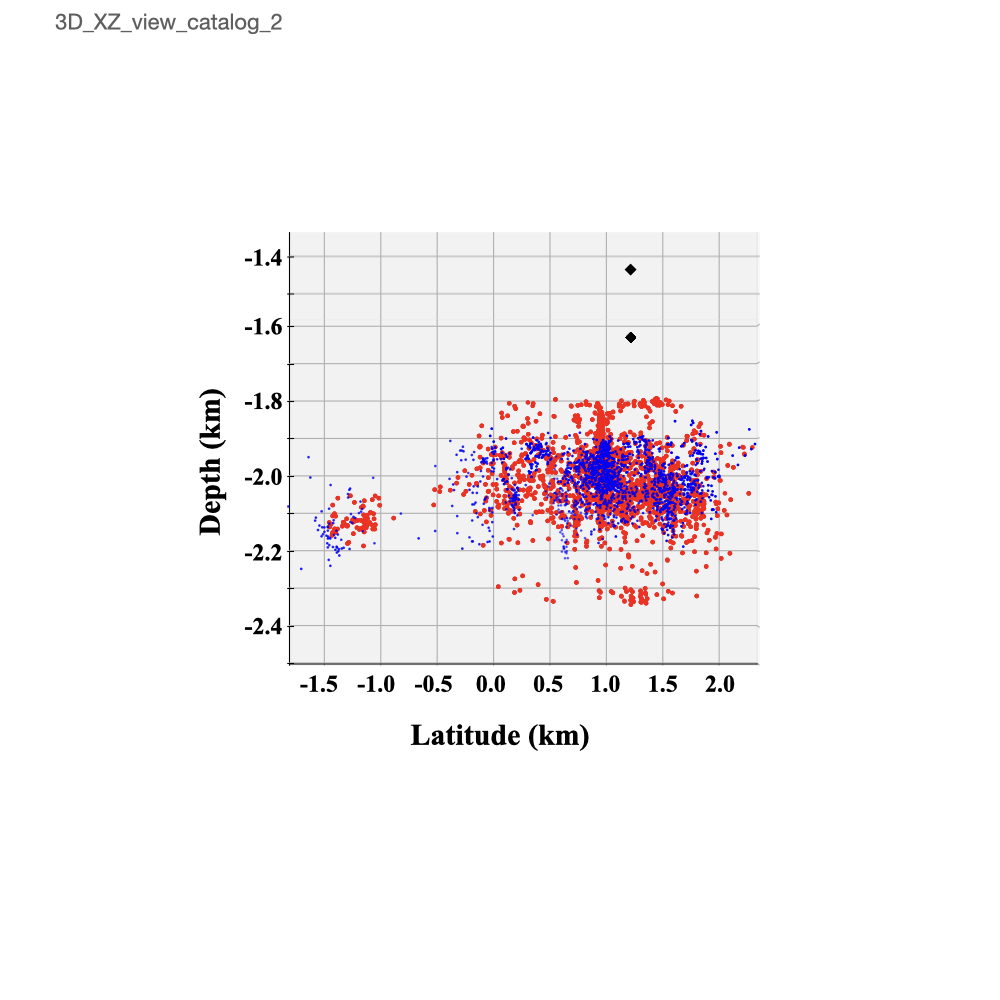}
    \caption{Latitude-depth view. }
    \label{fig:3D_XZ_view_catalog}
\end{subfigure}
\begin{subfigure}[b]{0.31\columnwidth}
    \centering
    \includegraphics[trim={7cm 8.5cm 7cm 6cm},clip,width=1\columnwidth]{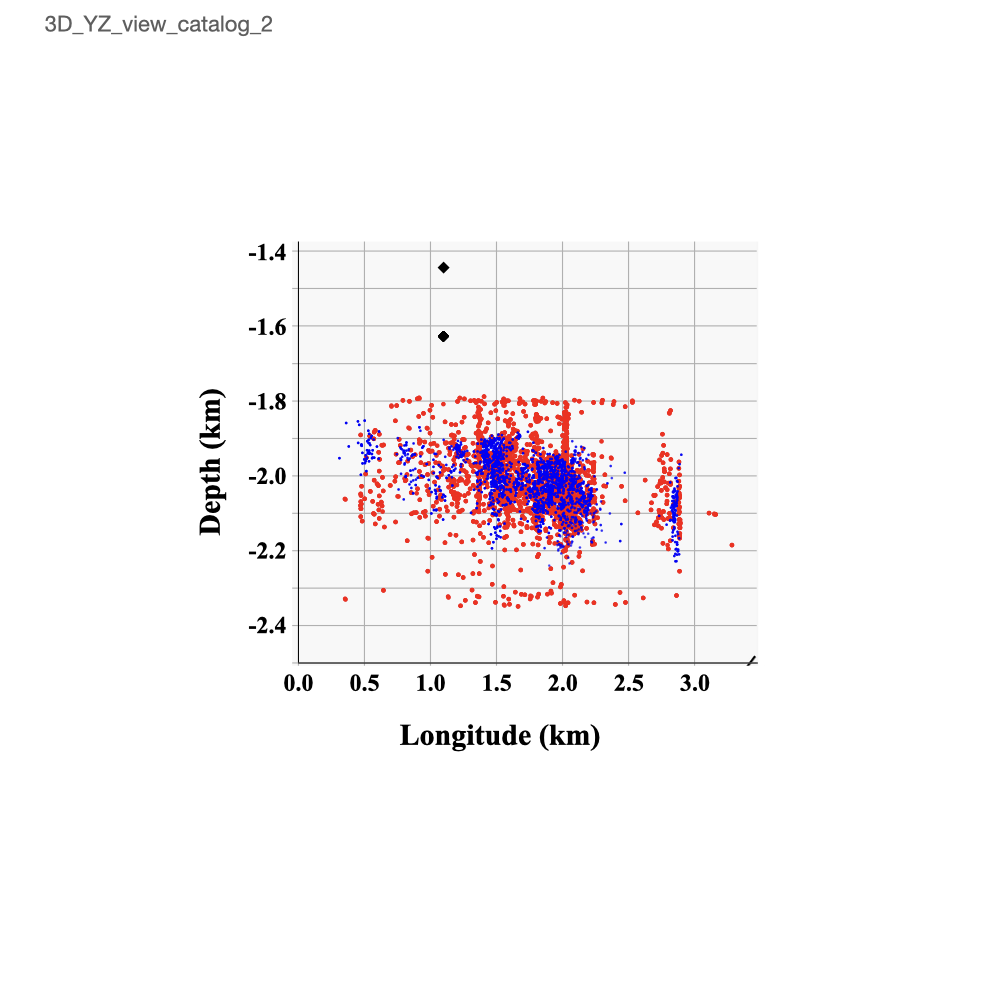}
    \caption{Longitude-depth view. }
    \label{fig:3D_YZ_view_catalog}
\end{subfigure}
\begin{subfigure}[b]{0.31\columnwidth}
    \centering
    \includegraphics[trim={6cm 6cm 7cm 9cm},clip,width=1\columnwidth]{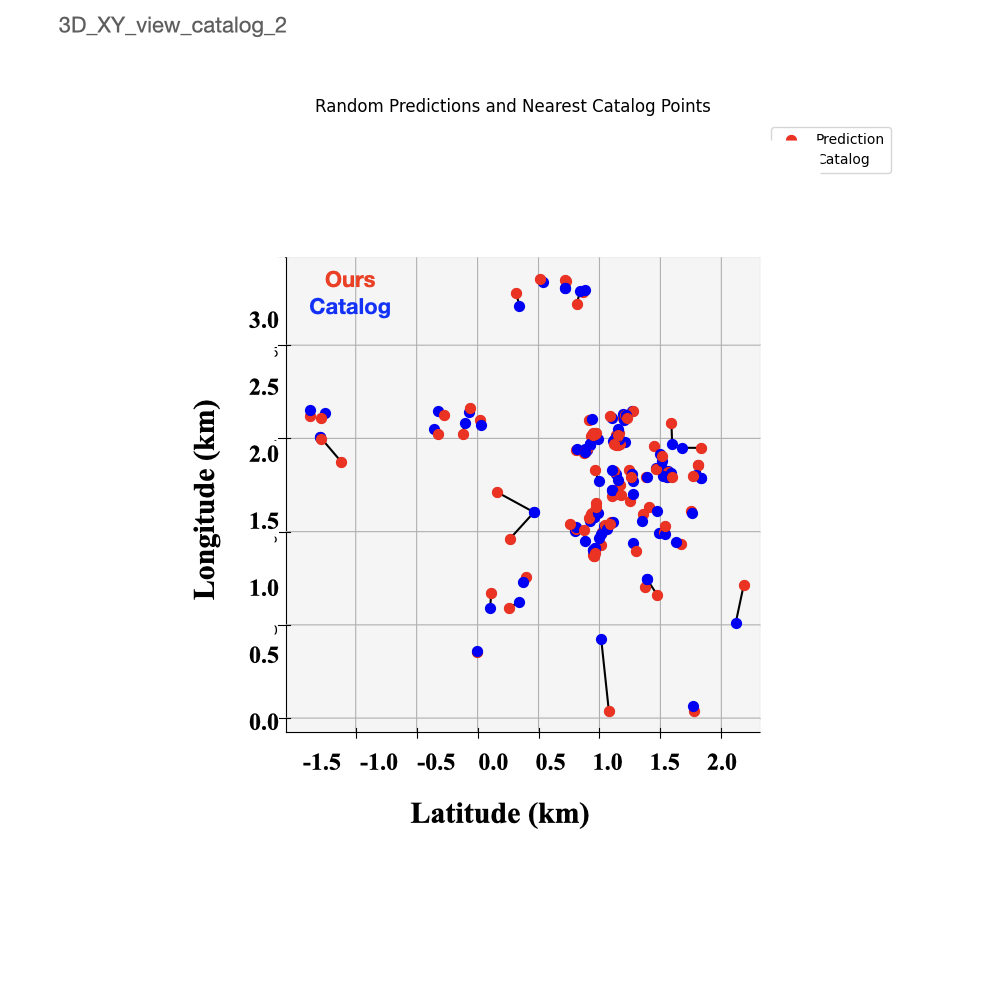}
    \caption{Longitude-latitude view. }
    \label{fig:3D_XY_view_catalog_compare}
\end{subfigure}
\begin{subfigure}[b]{0.31\columnwidth}
    \centering
    \includegraphics[trim={7cm 8.5cm 7cm 6cm},clip,width=1\columnwidth]{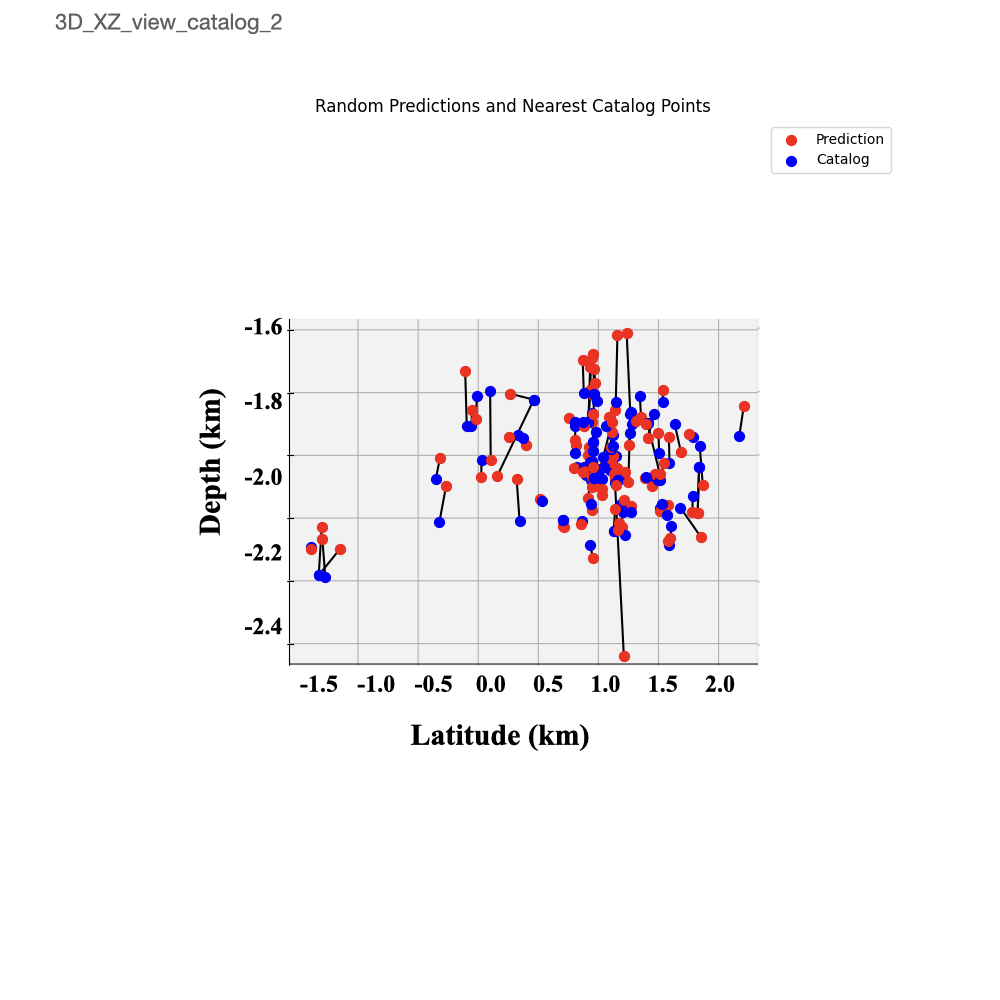}
    \caption{Latitude-depth view. }
    \label{fig:3D_XZ_view_catalog_compare}
\end{subfigure}
\begin{subfigure}[b]{0.31\columnwidth}
    \centering
    \includegraphics[trim={7cm 8.5cm 7cm 6cm},clip,width=1\columnwidth]{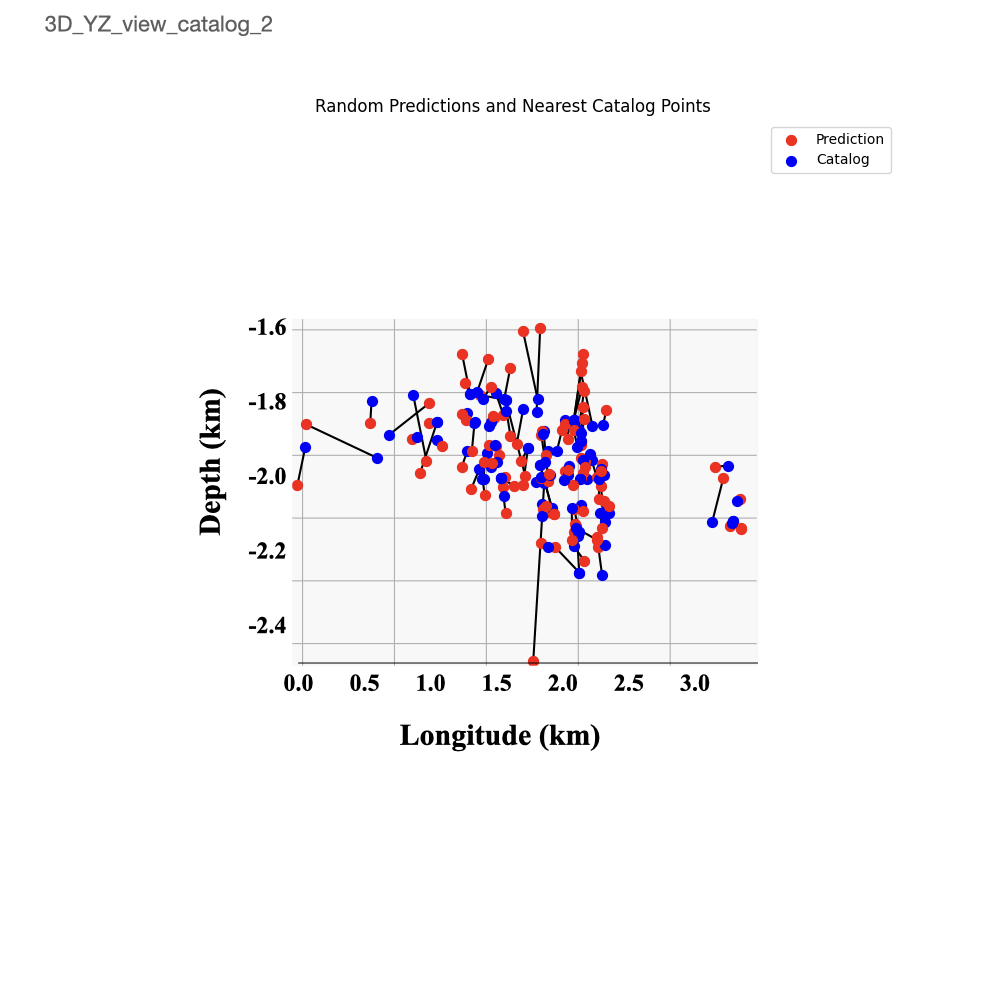}
    \caption{Longitude-depth view. }
    \label{fig:3D_YZ_view_catalog_compare}
\end{subfigure}
\caption{(Red) Catalog event locations calculated by the modified double-difference method. (Blue) Predicted event locations by the proposed method. The predicted locations are interpolated using the maximum and second maximum values of the heat-maps. The black diamonds are the CCS-1 injection ports. (a) Longitude-latitude, (b) latitude-depth, and (c) longitude-depth views. Comparison of 100 randomly selected events between the prediction and catalog locations are shown in (d) - (e), respectively. }
\label{fig:catalog}
\end{figure*}
When we compared our results with the previous modified double-difference method catalog locations~\cite{dando2021relocating}, shown in Figure~\ref{fig:catalog} as the blue and red dots, respectively, our algorithm successfully captures the events in most clusters. We calculate the median and mean of the hypocentral differences for the events between the catalog and our result. The median and mean equal to $0.04229$km and $0.06228$km, respectively. It suggests a reasonable match of our result to the catalog. However, some scattered events exist, especially between the main and smaller clusters. This variance might stem from poor data quality for certain events.

\begin{figure*}[!ht]
\centering
\begin{subfigure}[b]{0.45\columnwidth}
    \centering
    \includegraphics[trim={4cm 2cm 0cm 6cm},clip,width=1\columnwidth]{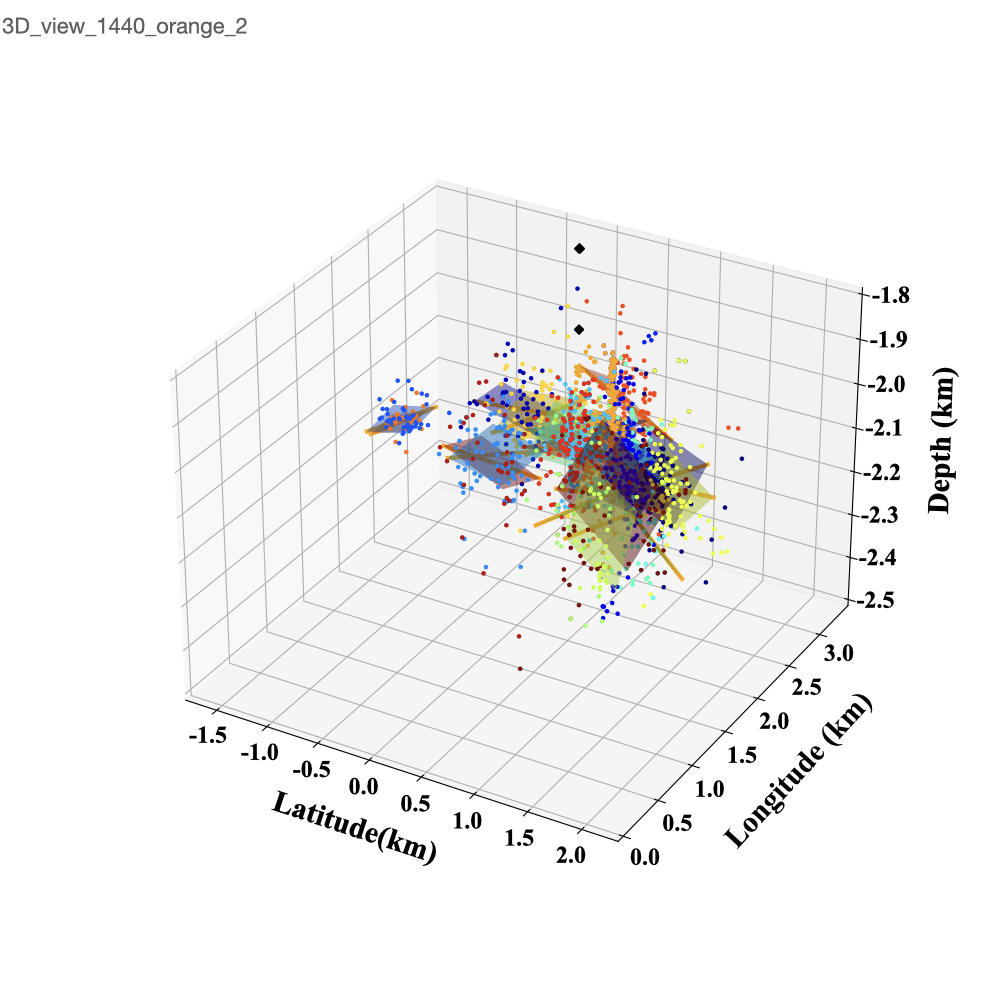}
    \caption{3D view}
    \label{fig:3D_view_1440}
\end{subfigure}
\begin{subfigure}[b]{0.45\columnwidth}
    \centering
    \includegraphics[trim={4cm 4cm 4cm 6cm},clip,width=1\columnwidth]{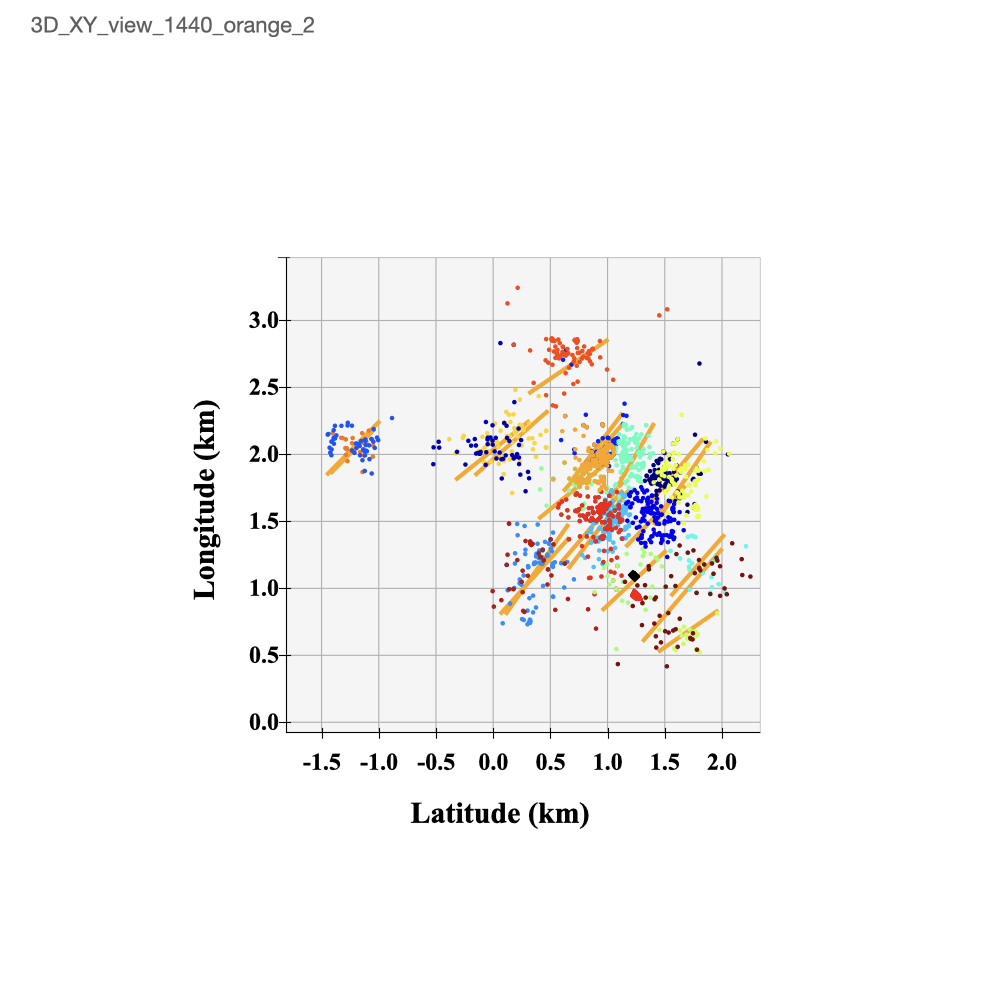}
    \caption{Longitude-latitude view}
    \label{fig:3D_XY_view_1440}
\end{subfigure}
\begin{subfigure}[b]{0.45\columnwidth}
    \centering
    \includegraphics[trim={4cm 6cm 4cm 6cm},clip,width=1\columnwidth]{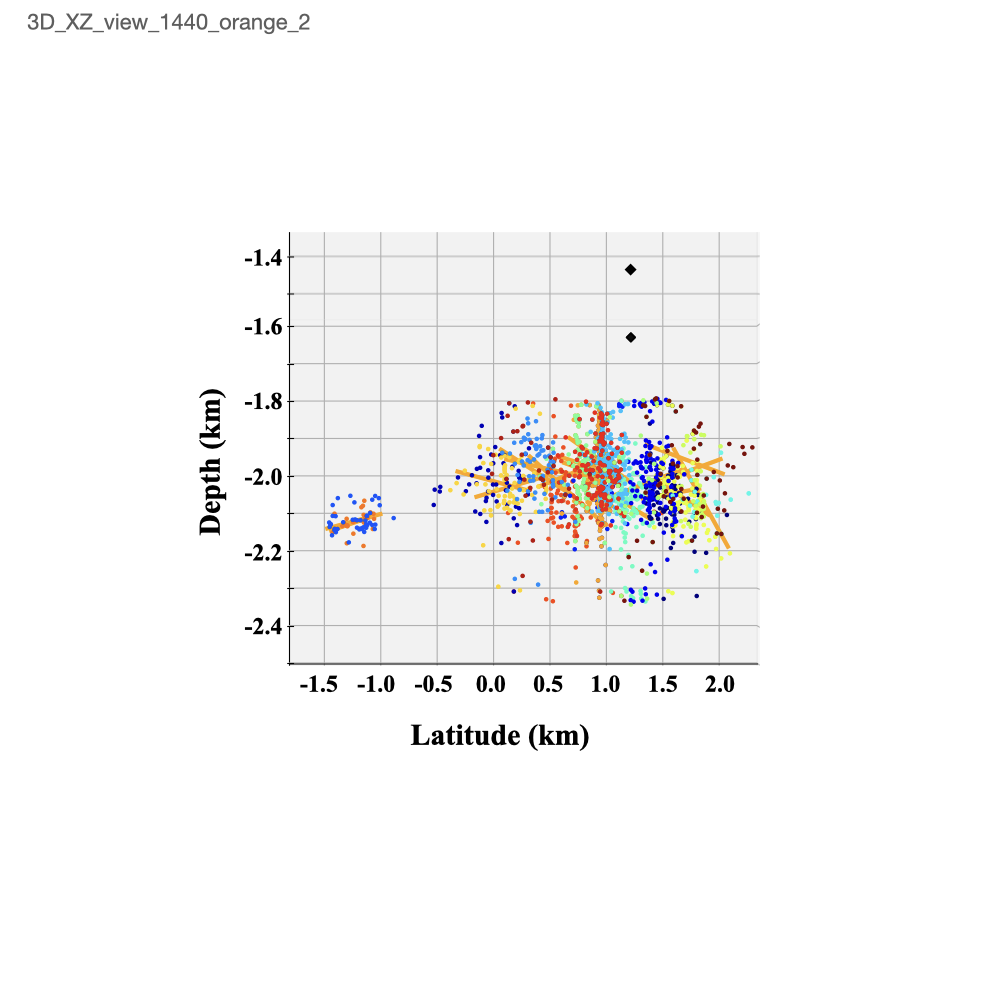}
    \caption{Latitude-depth view}
    \label{fig:3D_XZ_view_1440}
\end{subfigure}
\begin{subfigure}[b]{0.45\columnwidth}
    \centering
    \includegraphics[trim={4cm 6cm 4cm 6cm},clip,width=1\columnwidth]{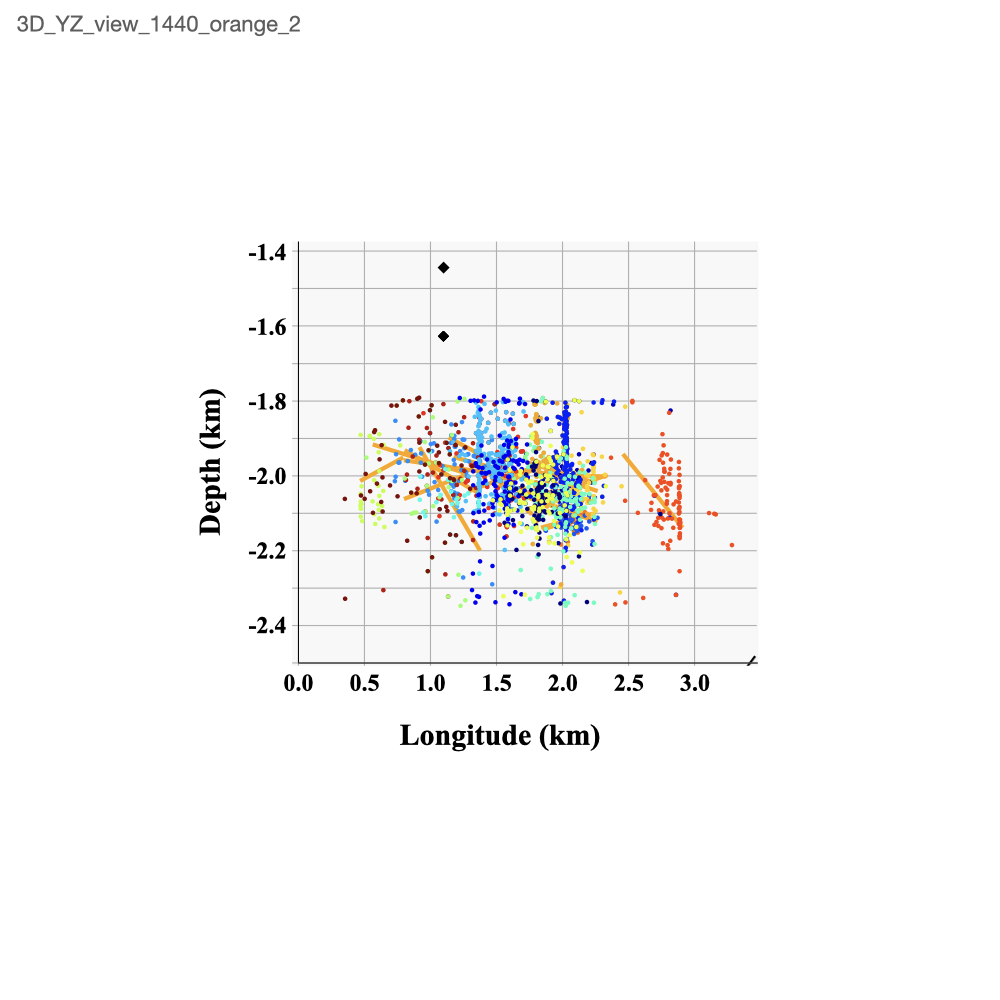}
    \caption{Longitude-depth view}
    \label{fig:3D_YZ_view_1440}
\end{subfigure}
\caption{Delineated faults and fractures for the k-mean clusters (orange lines) with a clustering time window of 1440 days. (a) 3D, (b) longitude-latitude, (c) latitude-depth, and (d) longitude-depth views. The locations are interpolated by the maximum and second maximum values of the heat-maps. The red triangle in (b) is the GM-1 borehole receivers. The black diamonds are the CCS-1 injection ports. Different colored dots are the k-mean clusters of the located events by the proposed method.}
\label{fig:3D_1440}
\end{figure*}

\begin{figure*}[!ht]
\centering
%\begin{subfigure}[b]{0.45\columnwidth}
%    \centering
%    \includegraphics[trim={4cm 2cm 0cm 6cm},clip,width=1\columnwidth]{./Fig/uncertainty_3D_2.png}
%    \caption{3D view}
%    \label{fig:3D_view_uq}
%\end{subfigure}
%\begin{subfigure}[b]{0.45\columnwidth}
%    \centering
    \includegraphics[trim={4cm 6cm 4cm 6cm},clip,width=0.75\columnwidth]{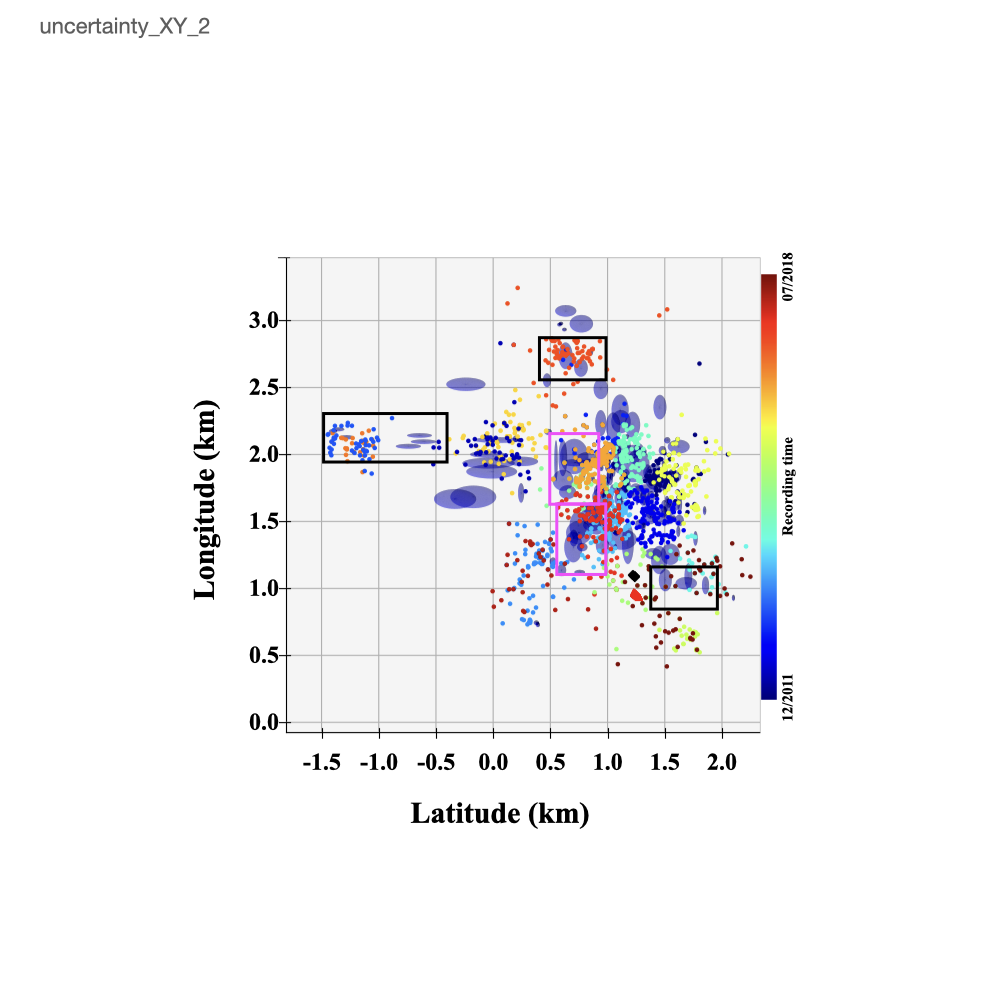}
%    \caption{Latitude-longitude view}
%    \label{fig:3D_XY_view_uq}
%\end{subfigure}
%\begin{subfigure}[b]{0.45\columnwidth}
%    \centering
%    \includegraphics[trim={4cm 6cm 4cm 6cm},clip,width=1\columnwidth]{./Fig/uncertainty_XZ_2.png}
%    \caption{Longitude-depth view}
%    \label{fig:3D_XZ_view_uq}
%\end{subfigure}
%\begin{subfigure}[b]{0.45\columnwidth}
%    \centering
%    \includegraphics[trim={4cm 6cm 4cm 6cm},clip,width=1\columnwidth]{./Fig/uncertainty_YZ_2.png}
%    \caption{Latitude-depth view}
%    \label{fig:3D_YZ_view_uq}
%\end{subfigure}
\caption{Dropout uncertainty analysis for the relocated seismic events on the Latitude-longitude plane. The location uncertainty ranges (blue spheres) are determined by $80\%$ cut-off Gaussian functions fitted by 50 independent dropout tests. Different colored dots are the k-mean clusters of the located events by the proposed method. The black boxes represent the far away clusters with smaller uncertainty; the magenta boxes represent the nearby clusters to the injection well with larger uncertainty.}
\label{fig:3D_uq}
\end{figure*}

With the relocated events, we further estimate the fault planes using the K-means clustering and the least distance fitting method as detailed in Section~\ref{sec:clustering}. Figure~\ref{fig:3D_view_1440} show cases the estimated fault planes in 3D, with different plane colors signifying fault triggering times. In Figures~\ref{fig:3D_XY_view_1440} to~\ref{fig:3D_YZ_view_1440}, we show the strike line of the fault planes as the solid orange lines. Our fault strike lines predominantly orient in a north-east to south-west direction, consistent with microseismic clusters observed during CO2 injection activities at the Decatur site, as documented by~\cite{williams2020analysis}. These authors noted variable reservoir responses to fluid pressure changes, which could explain the reactivation of adjacent faults in our study area. This aligns with their findings where similar fault behaviors were triggered by different injection rates and pressure changes, underscoring the complex interplay between geological settings and induced seismicity. Our algorithm saw them as separate because the temporal information difference between the two clusters at the same spatial region, but they might be a reactivation of the same fault.
To assess the robustness of our network performance, we modified the network architecture by introducing a dropout layer after each convolutional layer, except the final one, as described in Section~\ref{sec:uq}. The event uncertainty levels on an X-Y horizontal plane is illustrated in Figure~\ref{fig:3D_uq} as the blue ellipsoids. Generally speaking, events that have more unique features in the data and separated from the main cluster, marked inside the black boxes, were more accurately identified due to distinctive features like the larger moveout. The averaged uncertainty range inside the black boxes is within $0.15km$. However, events in close proximity, marked inside the magenta boxes, posed relatively higher uncertainty (up to $0.8km$) due to their similar features. Such uncertainty also accounts for the scattered event locations around the primary cluster. To address this, data from additional monitoring stations, like ISGS and USGS, should be considered in future analyses.
\section{Discussion and Limitations}\label{sec:discussion}

\subsection{Short-term monitoring of fracture delineation}
\begin{figure*}[!ht]
\centering
\begin{subfigure}[b]{0.45\columnwidth}
    \centering
    \includegraphics[trim={4cm 2cm 0cm 6cm},clip,width=1\columnwidth]{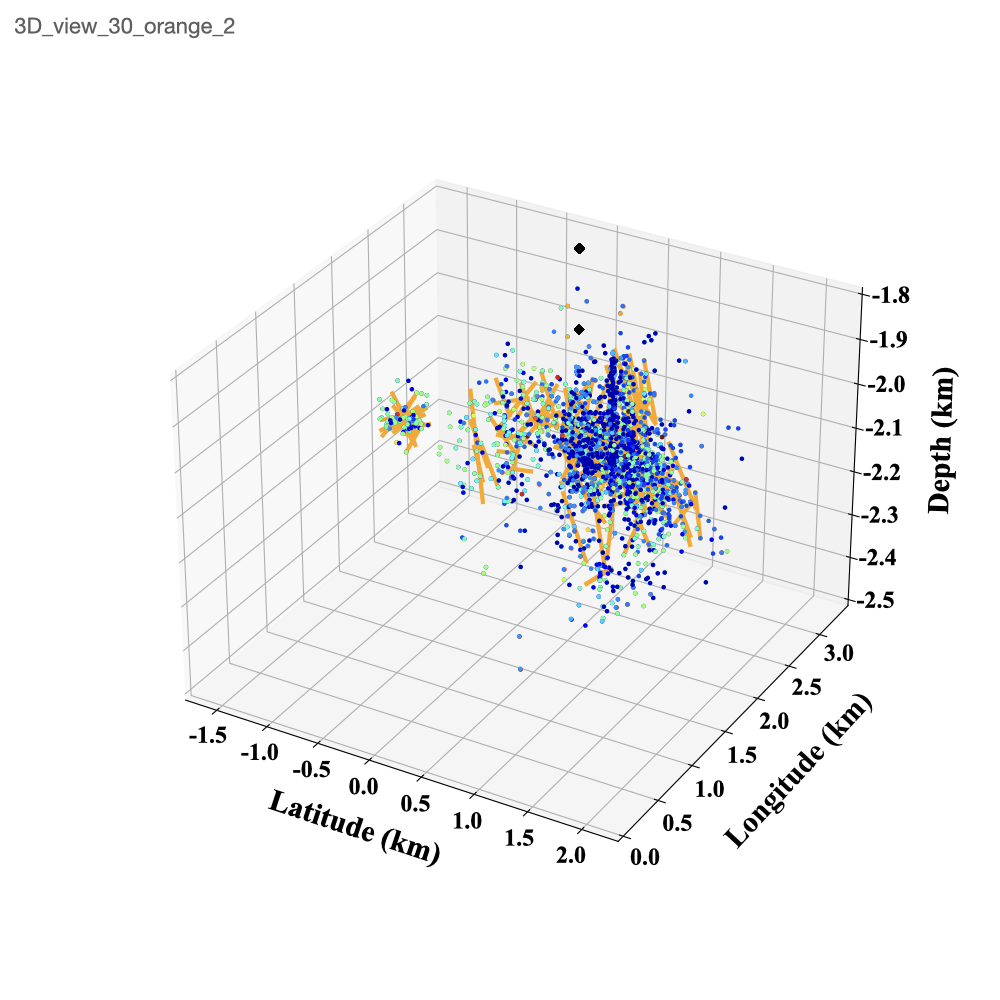}
    \caption{3D view}
    \label{fig:3D_view_30}
\end{subfigure}
\begin{subfigure}[b]{0.45\columnwidth}
    \centering
    \includegraphics[trim={4cm 4cm 4cm 6cm},clip,width=1\columnwidth]{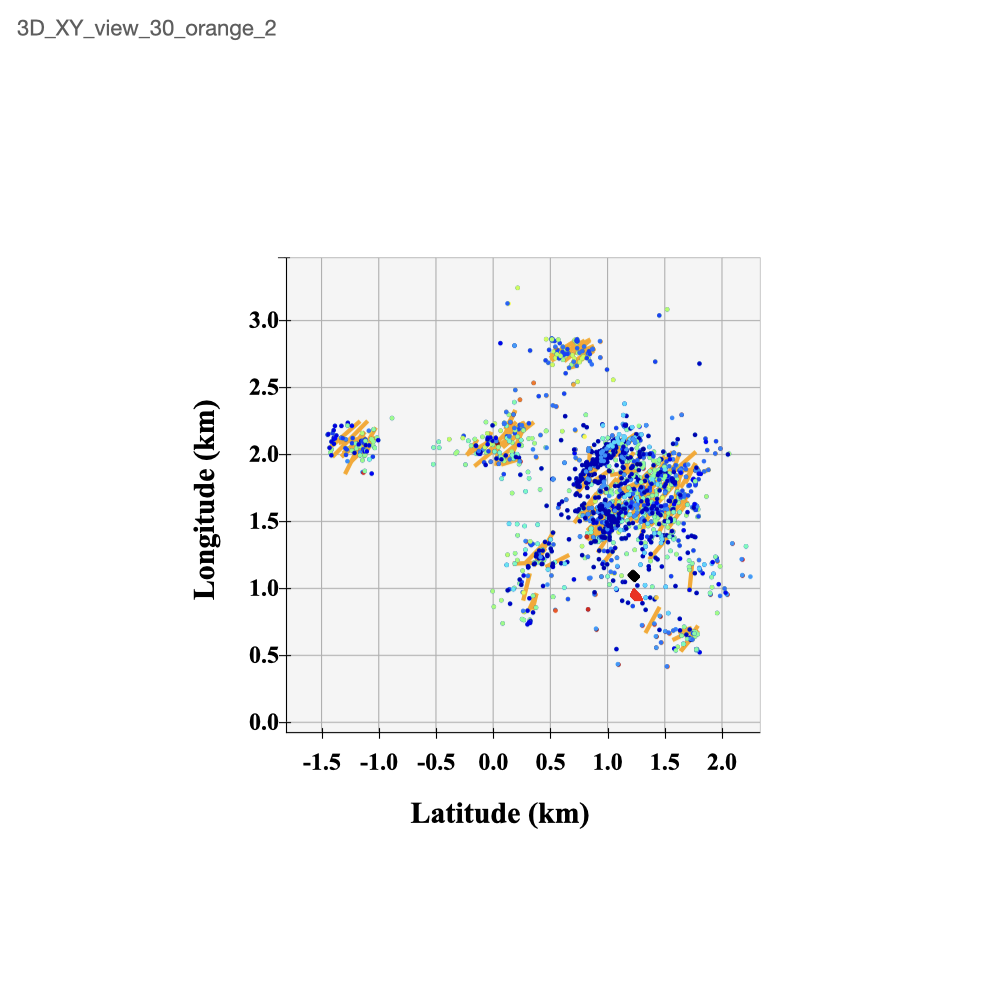}
    \caption{Longitude-latitude view}
    \label{fig:3D_XY_view_30}
\end{subfigure}
\caption{Short-term delineated faults and fractures for the k-mean clusters (orange lines) with a clustering time window of 30 days. (a) 3D, (b) longitude-latitude views. The locations are interpolated by the maximum and second maximum values of the heat-maps. The red triangles are the GM-1 borehole receivers. The black diamonds are the CCS-1 injection ports. Different colored dots are the k-mean clusters of the located events in different temporal period by the proposed method.}
\label{fig:3D_30}
\end{figure*}

Besides the long-term (1440 days) fault delineation showed in the previous section, we also performed a short-term fracture delineation using the same algorithm, except for the temporal period segments, which are set to 30 days. The results can be found in Figure~\ref{fig:3D_30}. The fault lines in the short-term plots suggest that the regional faults and fractures are re-activated multiple times instead of experiencing a one-time activation. On the other hand, the faults and fractures show almost the same distribution in space compared with the long-term delineation in Figures~\ref{fig:3D_XY_view_1440}. 
\subsection{Distance threshold strategy with iterative elimination}
\begin{figure*}[!ht]
\centering
\begin{subfigure}[b]{0.31\columnwidth}
    \centering
    \includegraphics[trim={6cm 6cm 7cm 9cm},clip,width=1\columnwidth]{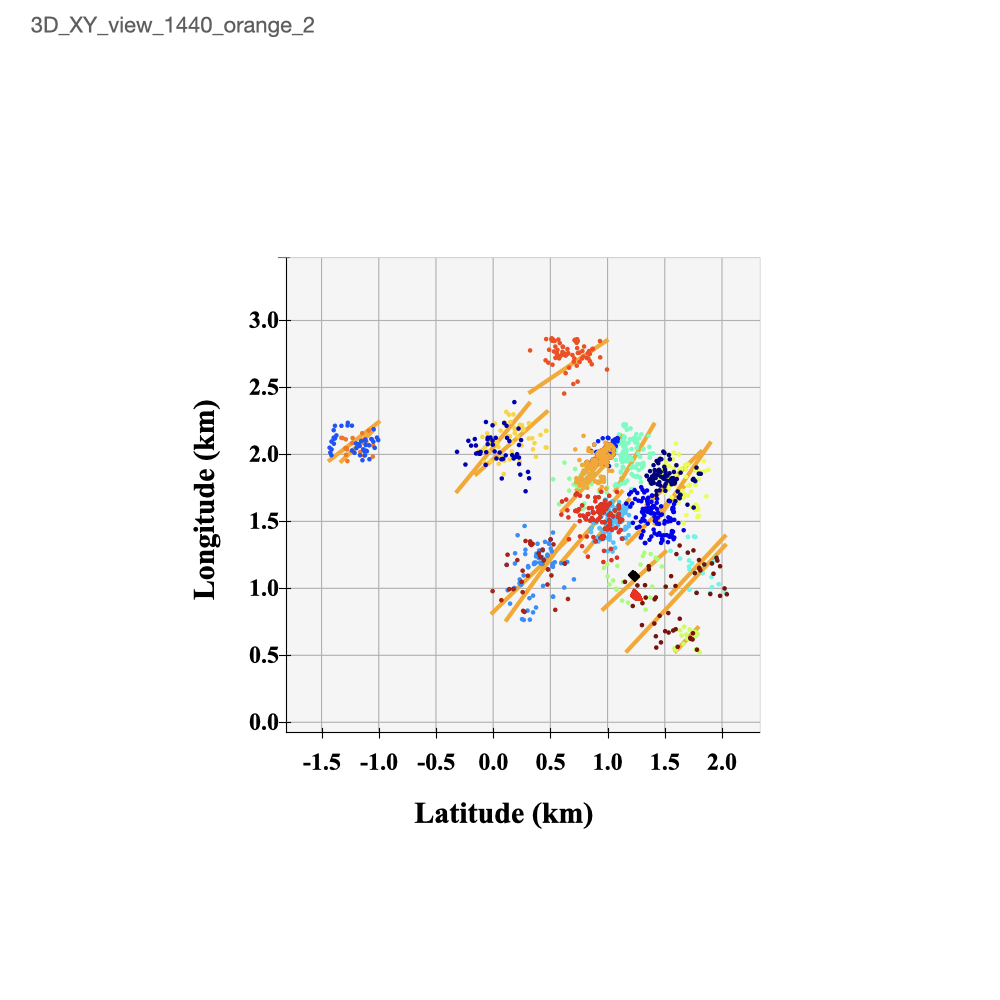}
    \caption{Longitude-latitude view. }
    \label{fig:iteration_XY}
\end{subfigure}
\begin{subfigure}[b]{0.31\columnwidth}
    \centering
    \includegraphics[trim={7cm 8.5cm 7cm 6cm},clip,width=1\columnwidth]{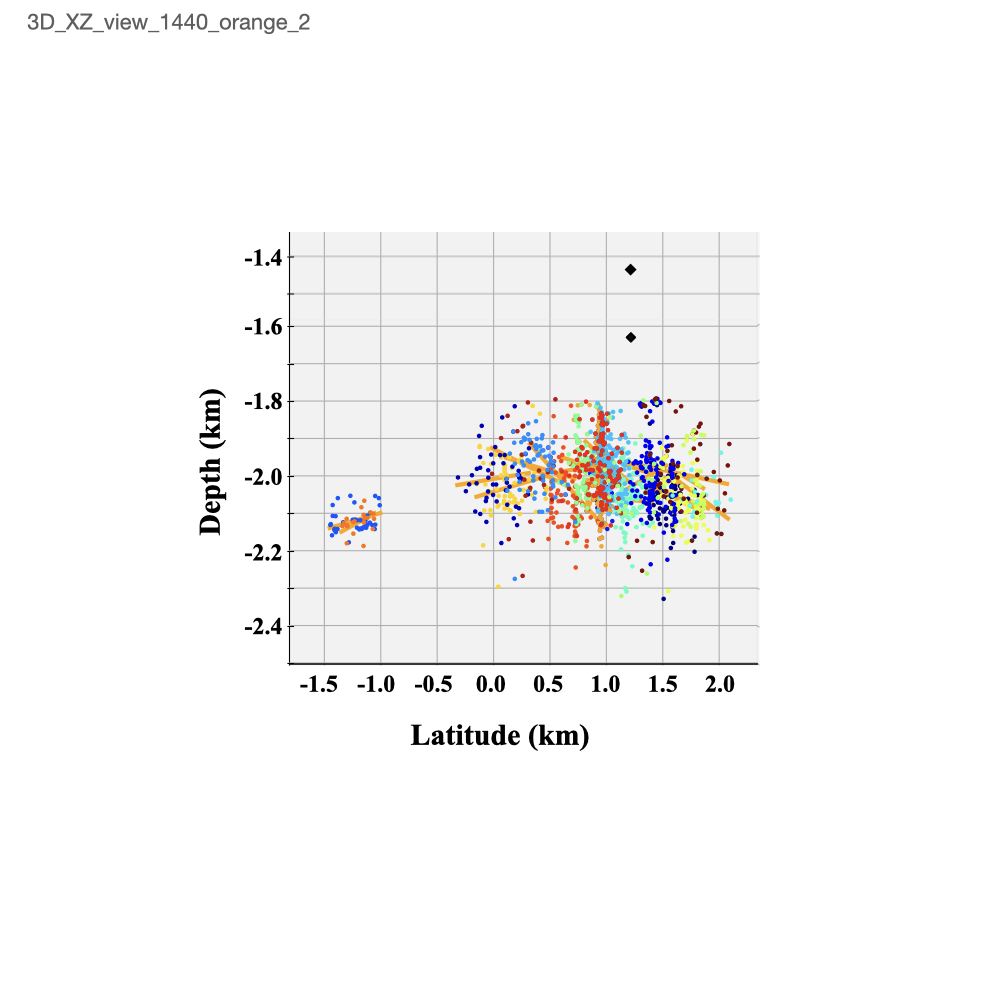}
    \caption{Latitude-depth view. }
    \label{fig:iteration_XZ}
\end{subfigure}
\begin{subfigure}[b]{0.31\columnwidth}
    \centering
    \includegraphics[trim={7cm 8.5cm 7cm 6cm},clip,width=1\columnwidth]{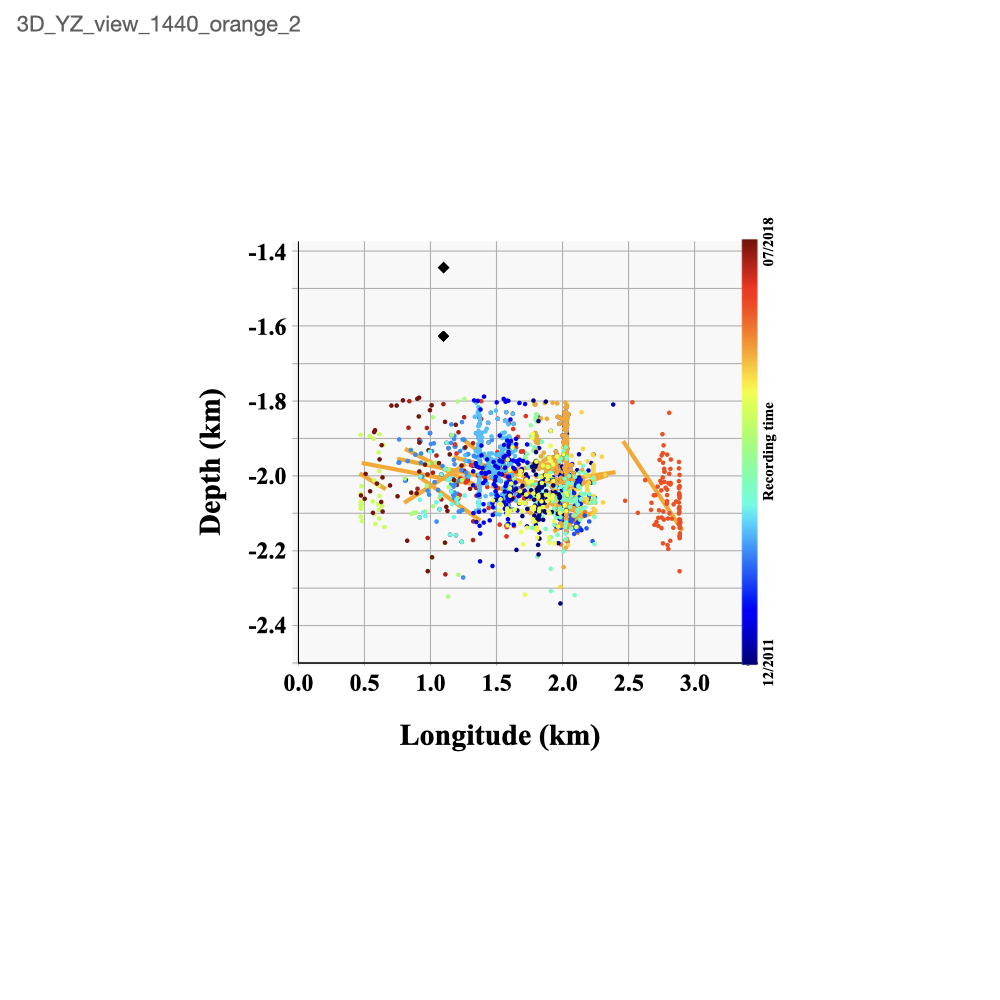}
    \caption{Longitude-depth view. }
    \label{fig:iteration_YZ}
\end{subfigure}
\caption{Fault line fitting results and the corresponding event locations with an iterative elimination method in the (a) longitude-latitude, (b) latitude-depth, and (c) longitude-depth views. The time window is 1440 days.}
\label{fig:iteration}
\end{figure*}

We set a fixed distance threshold in the fault plane fitting step to obtain the results in Figures~\ref{fig:3D_1440} and~\ref{fig:3D_30}. In this section, we discuss an alternative strategy of choosing the distance threshold. We use an iterative approach, where we first estimate the fault plane with all available events in the same cluster, and then only keep the events that are within $90$ percentile of all event distances to the fault plane. We repeat the two steps for 10 iterations. The iterative method generally eliminates the events that are too far away from the estimated fault plane. The results of the iterative method are shown in the Figure~\ref{fig:iteration}. The faults (orange lines) in the figure actually are very close to the original results in Figure~\ref{fig:3D_1440}, including similar fault locations, directions and dipping angles. It suggests that either threshold strategy works well and one method proves the accuracy of the other.
\subsection{Distance threshold strategy with RANSAC}
\begin{figure*}[!ht]
\centering
\begin{subfigure}[b]{0.31\columnwidth}
    \centering
    \includegraphics[trim={6cm 6cm 7cm 9cm},clip,width=1\columnwidth]{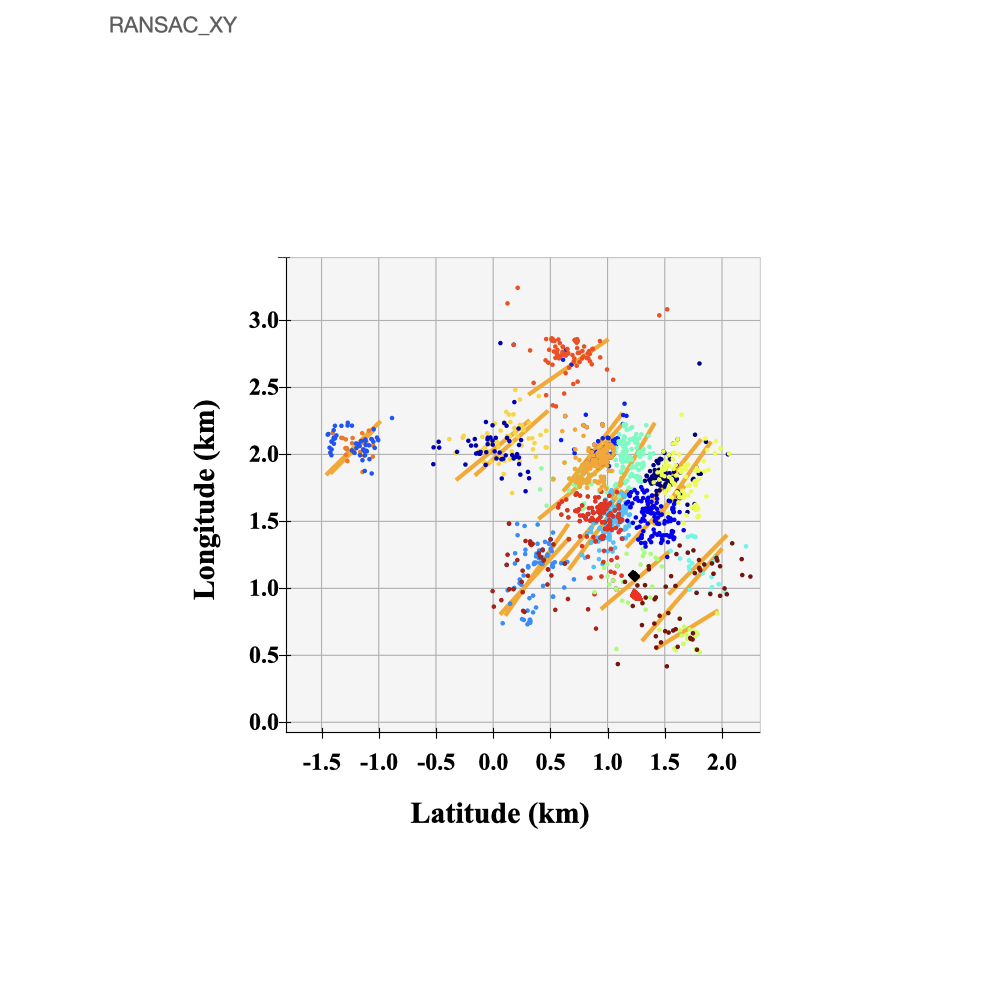}
    \caption{Longitude-latitude view. }
    \label{fig:ransac_XY}
\end{subfigure}
\begin{subfigure}[b]{0.31\columnwidth}
    \centering
    \includegraphics[trim={7cm 8.5cm 7cm 6cm},clip,width=1\columnwidth]{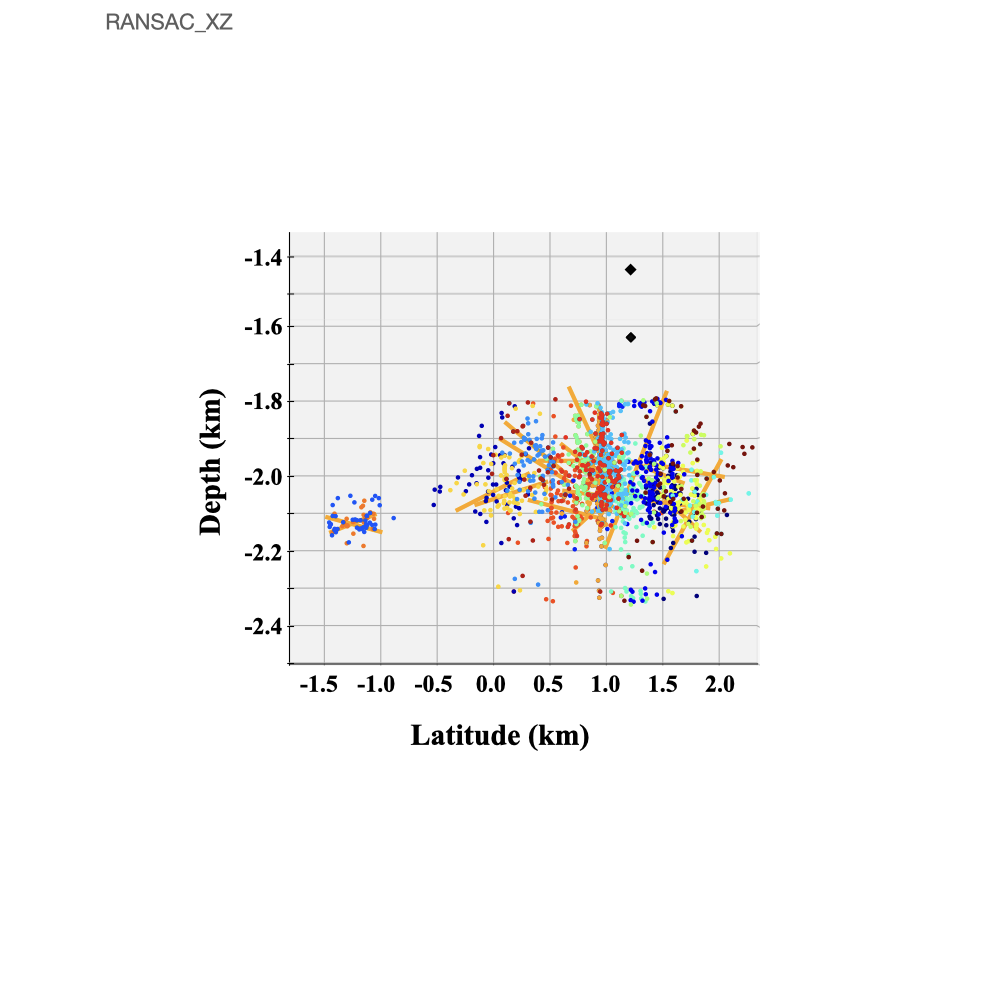}
    \caption{Latitude-depth view. }
    \label{fig:ransac_XZ}
\end{subfigure}
\begin{subfigure}[b]{0.31\columnwidth}
    \centering
    \includegraphics[trim={7cm 8.5cm 7cm 6cm},clip,width=1\columnwidth]{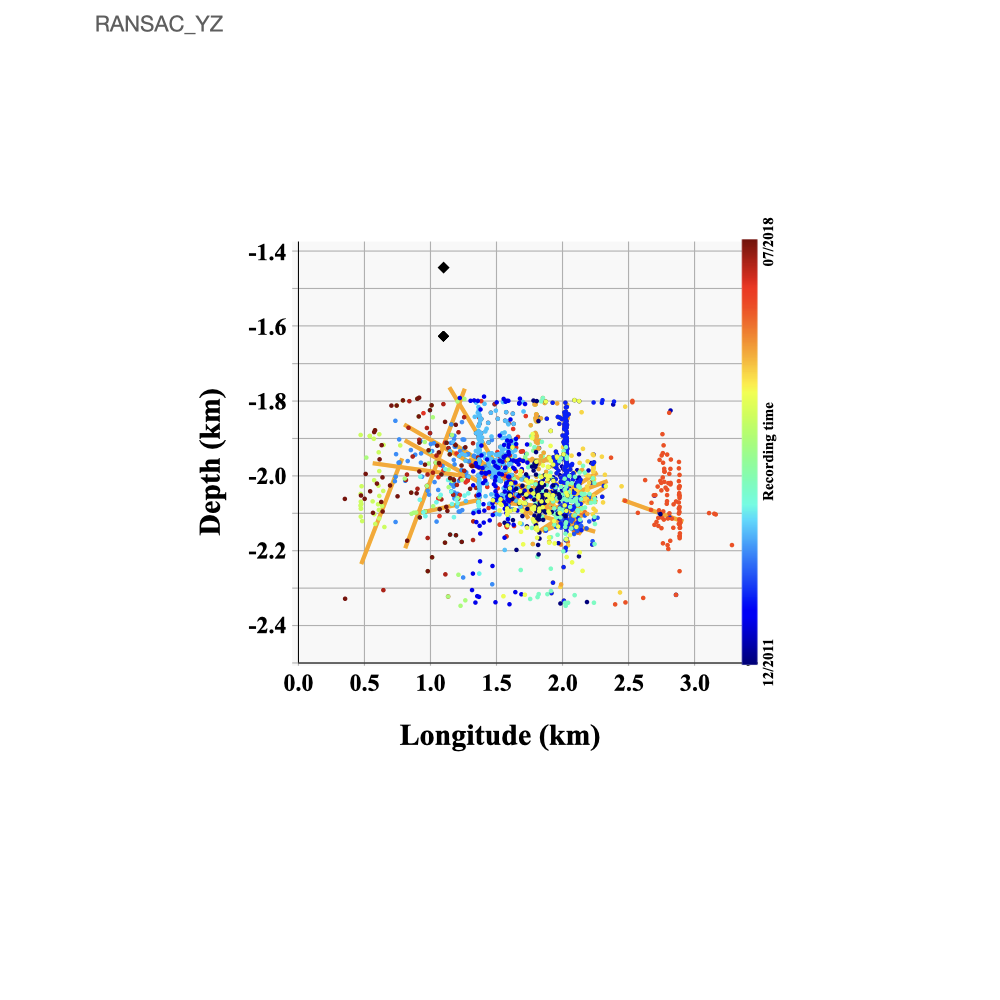}
    \caption{Longitude-depth view. }
    \label{fig:ransac_YZ}
\end{subfigure}
\caption{Fault line fitting results and the corresponding event locations with the RANSAC algorithm in the (a) longitude-latitude, (b) latitude-depth, and (c) longitude-depth views. The time window is 1440 days.}
\label{fig:ransac}
\end{figure*}

To improve the robustness of our fault plane fitting, we adopted the Random Sample Consensus (RANSAC) algorithm, which inherently handles outliers more effectively than a fixed distance threshold. In contrast to the iterative method discussed previously, RANSAC does not require a predefined distance threshold. Instead, it iteratively selects a subset of the total data set, fits a model, and determines the number of inliers that fall within a certain distance from the model. This process repeats, and the model with the greatest number of inliers is selected as the best fit. The RANSAC algorithm has been successfully applied to estimating seismogenic fault structures~\cite{skoumal2019characterizing}, and thus, we apply the algorithm in our study to show the robustness of the DeFault algorithm. Figures~\ref{fig:ransac_XY}, ~\ref{fig:ransac_XZ}, and ~\ref{fig:ransac_YZ} show the fault line fitting results and the event locations used in the RANSAC approach. The time window for these results is 1440 days. Interestingly, the RANSAC results are quite consistent with those obtained using previous methods such as K-means clustering and iterative elimination. Particularly in the longitude-latitude (XY) plane view, the fault lines identified by RANSAC closely match those derived from earlier approaches. This consistency across different methods not only confirms the robustness of our fault plane estimation but also provides a statistically sound basis for rejecting outliers, thereby improving the reliability of our geological interpretations.

\subsection{Computational cost}
The proposed $\mathbf{\emph{DeFault}}$ method is very efficient in the inference stage, where the raw recorded seismic events can be relocated and clustered within 4 seconds per seismic data on single V100 NVIDIA GPU, including the signal enhancement process, relocation and fault fitting steps. This computational cost is almost negligible when compared to the traditional physics-driven methods because $\mathbf{\emph{DeFault}}$ is purely data-driven and does not require wavefield extrapolation, such as waveform inversion. In the training stage, since the network training heavily relies on the training set, the 3D acoustic simulation of the synthetic training data is computationally costly, requiring around 30 minutes per sample on a single CPU. However, note that the training samples need to be simulated only once. The network was trained on a twin-GPU HPC cluster node with both GPUs given by Nvidia A100. The total training time cost is twelve hours. The one-time cost of synthetic simulation and the training cost are affordable when labelled field data are not available. However, if higher spatial resolution is required, for example to reduce the grid effect of nearby events, larger heat-maps are necessary for training, which may increase the computational cost by the order of resolution difference. 

\subsection{Limitations}
Our proposed methodology attempts to bridge the gap between synthetic and real data distributions by leveraging the \textit{MLReal transformations} on both source and target datasets. Despite their merit, the approach is not without constraints.

A primary drawback emerges from the assumption that the vertical axis in input sections — whether images or shot gathers — doesn't significantly influence the absolute values pertinent to the task. While this assumption paves the way for the domain adaptation transformation, it might not universally apply. For tasks dependent on the accuracy of the input's vertical axis' absolute values, this assumption could introduce errors. However, in our specific application, the vertical (time) axis is less important as we rely on the moveout information of the events, which are well-kept after the correlation and convolution operations. 

Additionally, the transformations employed in this strategy predominantly utilize linear operations. Even though these operations work well to match the differences in energy, noise, and other attributes between synthetic and real datasets, certain applications might need intricate transformations. In such scenarios, delving into more advanced techniques, including deep learning-driven domain adaptation methodologies, could be beneficial.

Moreover, the simulations of synthetic training samples are highly dependent on the precision of the existing velocity model. As such, any discrepancies in the velocity model could distort the training sample accuracy, thereby accumulating both training and testing errors of the neural network. Besides, we ignored the elastic effect in our study, and we anticipated that our approach can be further improved by using elastic simulation of the seismic events and incorporating both P and S wave arrivals in the relocation step.

Yet, despite these constraints, our proposed framework utilizes a potent tool for domain adaptation in supervised learning, especially when the target dataset lacks labels. By aligning the distributions of both synthetic and real data, our approach augments the neural network models' capability to generalize when interfacing with field data, thus enhancing performance across diverse operations. An added advantage lies in its computational prowess during the inference phase—it boasts a swift processing time of roughly 4 seconds per sample after event detection. This briskness holds promise for real-time seismic monitoring and fault delineating, leading to an era of immediate CO$_2$ injection hazard monitoring.

By sharing our findings and innovative methods with the academic community and CCUS stakeholders, we hope to foster collaborative efforts and promote knowledge exchange. It's crucial that as CCUS technology matures and garners broader acceptance, researchers and industry specialists work hand-in-hand to ensure the highest safety standards are maintained. Such collective dedication will undoubtedly fortify global initiatives to tackle climate change, propelling us towards a more environmentally-responsible future.

\subsection{Scope of the Study}
This study serves as a preliminary exploration of machine learning-enhanced passive seismic monitoring, demonstrating the potential of our algorithm across various applications, including hydraulic fracturing, geothermal activities, and volcano monitoring. While our experiments focused on the Illinois Basin-Decatur Project (IBDP) due to funding availability and data accessibility, the algorithm's utility is not confined to carbon capture, utilization, and storage (CCUS) applications alone. We have addressed a basic scenario in seismic event relocation and fault imaging, where filters were predefined to optimize training efficacy. It’s important to note that the IBDP data, like many seismic monitoring datasets, typically feature events centered temporally. Once filters are tailored for specific seismic sites, they remain fixed across both training and subsequent field data applications, as the MLReal steps ensure consistent data distribution. However, the current implementation may obscure source-mechanism information due to MLReal and amplitude normalization processes, and relies solely on Vp models in simulations. This simplification limits the algorithm's applicability to more complex inversion tasks, suggesting the need for future research to refine these methods and extend their capabilities.

\section{Conclusion}\label{sec:conclusion}

Our research contributes to the recognized need for enhanced fault delineation and analysis in CCUS initiatives by presenting a novel approach that facilitates these processes. Through the introduction and validation of the $\mathbf{\emph{DeFault}}$ methodology, we aimed to bridge the gap between field passive seismic monitoring data and synthetic simulations, offering a valuable resource for the IBDP project CCUS monitoring task. While our innovation demonstrates a potential for increasing the precision of fault identification in CO$_2$ reservoir regions, its most significant contribution lies in dramatically reducing the time needed to obtain results. This expedited process could be particularly valuable for operators making critical decisions, where the immediacy of information often takes precedence over absolute precision.

Using the \textit{MLReal} technique, $\mathbf{\emph{DeFault}}$ harmoniously combines synthetic and field data into a unified distribution. This enables the training of our encoder-decoder—designed to extract moveout features from the input seismic data and provide the corresponding source location information—without relying on labeled field data. Equipped with relocated passive sources, we've delineated faults and fractures using the K-means algorithm and the least distance fitting method. Our algorithm is versatile, suitable for both short-term and long-term delineation tasks in the IBDP CCUS project.

Our assessment of the $\mathbf{\emph{DeFault}}$ method, applied to the Decatur CO$_2$ storage project, has yielded valuable insights. By utilizing the publicly available IBDP data and concentrating on well data, we've harnessed its clearest features. Our three-dimensional approach generated a synthetic training set comprising 15,000 samples, tailored to match the IBDP data's receiver geometry. Post signal processing, the integration of processed field and synthetic data was executed using the MLReal technique. Testing our trained neural network against field data produced predictions largely aligned with previous studies. However, certain anomalies were detected, likely attributed to data quality inconsistencies. Our in-depth analysis of fault planes revealed patterns that align with prior research outcomes, especially with respect to identifying reactivated faults. This alignment suggests our methodology's findings are comparable to existing studies, offering further validation of our approach. In assessing the accuracy of our predictions, we discovered a higher precision in detecting events farther from sensors, while closer events presented challenges—highlighting potential avenues for future refinement, like incorporating data from additional monitoring stations.

\section*{Open Research Section}
The IBDP data used in this study~\cite{illinois_netl} can be retrieved on the NETL's EDX platform at~\url{https://doi.org/10.18141/1854142}. The code designed for $\mathbf{\emph{DeFault}}$ will be released within year 2024, upon approval by Los Alamos National Laboratory and Department of Energy of the United States. 

%Additionally, we showed that training an existing FWI model, such as InversionNet, on \textsc{WaveDiffusion}-generated samples improves performance, particularly when small amounts of real data are available. This underscores the potential of using generative models to supplement real datasets in seismic inversion tasks.

\section*{Acknowledgments}
We thank the SMART project by the Department of Energy of the United States for financially support this research. Research presented in this article was supported by the Laboratory Directed Research and Development program of Los Alamos National Laboratory under project number 20240322ER. We thank the EES-16 group of Los Alamos National Laboratory for the helpful discussion during the study.
%Bibliography
\bibliographystyle{unsrt}  
\bibliography{references}

\begin{thebibliography}{10}

\bibitem{pacala2004stabilization}
Stephen Pacala and Robert Socolow.
\newblock Stabilization wedges: solving the climate problem for the next 50 years with current technologies.
\newblock {\em science}, 305(5686):968--972, 2004.

\bibitem{white2016assessing}
Joshua~A White and William Foxall.
\newblock Assessing induced seismicity risk at co2 storage projects: Recent progress and remaining challenges.
\newblock {\em International Journal of Greenhouse Gas Control}, 49:413--424, 2016.

\bibitem{bauer2019}
Robert~A Bauer, Robert Will, Sallie~E Greenberg, and Steven~G Whittaker.
\newblock {\em Illinois Basin–Decatur Project}, pages 349--370.
\newblock Cambridge University Press, 2019.

\bibitem{finley2014}
R.J. Finley.
\newblock An overview of the illinois basin–decatur project.
\newblock {\em Greenhouse Gases: Science and Technology}, 4(5):571--579, 2014.

\bibitem{vasco2008reservoir}
DW~Vasco, Alessandro Ferretti, and Fabrizio Novali.
\newblock Reservoir monitoring and characterization using satellite geodetic data: Interferometric synthetic aperture radar observations from the krechba field, algeria.
\newblock {\em Geophysics}, 73(6):WA113--WA122, 2008.

\bibitem{JIANG2021110521}
Kai Jiang and Peta Ashworth.
\newblock The development of carbon capture utilization and storage (ccus) research in china: A bibliometric perspective.
\newblock {\em Renewable and Sustainable Energy Reviews}, 138:110521, 2021.

\bibitem{TAPIA20181}
John Frederick~D. Tapia, Jui-Yuan Lee, Raymond~E.H. Ooi, Dominic~C.Y. Foo, and Raymond~R. Tan.
\newblock A review of optimization and decision-making models for the planning of co2 capture, utilization and storage (ccus) systems.
\newblock {\em Sustainable Production and Consumption}, 13:1--15, 2018.

\bibitem{HASAN20152}
M.M.~Faruque Hasan, Eric~L. First, Fani Boukouvala, and Christodoulos~A. Floudas.
\newblock A multi-scale framework for co2 capture, utilization, and sequestration: Ccus and ccu.
\newblock {\em Computers \& Chemical Engineering}, 81:2--21, 2015.
\newblock Special Issue: Selected papers from the 8th International Symposium on the Foundations of Computer-Aided Process Design (FOCAPD 2014), July 13-17, 2014, Cle Elum, Washington, USA.

\bibitem{waldhauser2000double}
F.~Waldhauser and W.~L. Ellsworth.
\newblock A double-difference earthquake location algorithm: Method and application to the northern hayward fault, california.
\newblock {\em Bulletin of the Seismological Society of America}, 90:1353--1368, 2000.

\bibitem{zhang2003double}
Haijiang Zhang and Clifford~H Thurber.
\newblock Double-difference tomography: The method and its application to the hayward fault, california.
\newblock {\em Bulletin of the Seismological Society of America}, 93(5):1875--1889, 2003.

\bibitem{kamei2014passive}
Rie Kamei and David Lumley.
\newblock Passive seismic imaging and velocity inversion using full wavefield methods.
\newblock In {\em SEG International Exposition and Annual Meeting}, pages SEG--2014. SEG, 2014.

\bibitem{sun2016full}
Junzhe Sun, Zhiguang Xue, Tieyuan Zhu, Sergey Fomel, and Norimitsu Nakata.
\newblock Full-waveform inversion of passive seismic data for sources and velocities.
\newblock In {\em SEG International Exposition and Annual Meeting}, pages SEG--2016. SEG, 2016.

\bibitem{wang2018microseismic}
Hanchen Wang and Tariq Alkhalifah.
\newblock Microseismic imaging using a source function independent full waveform inversion method.
\newblock {\em Geophysical Journal International}, 214(1):46--57, 2018.

\bibitem{8883032}
Chao Song and Tariq Alkhalifah.
\newblock Microseismic event estimation based on an efficient wavefield inversion.
\newblock {\em IEEE Journal of Selected Topics in Applied Earth Observations and Remote Sensing}, 12(11):4664--4671, 2019.

\bibitem{wang2020regularized}
Hanchen Wang, Qiang Guo, Tariq Alkhalifah, and Zedong Wu.
\newblock Regularized elastic passive equivalent source inversion with full-waveform inversion: Application to a field monitoring microseismic data set.
\newblock {\em Geophysics}, 85(6):KS207--KS219, 2020.

\bibitem{artman2010source}
Brad Artman, Igor Podladtchikov, and Ben Witten.
\newblock Source location using time-reverse imaging.
\newblock {\em Geophysical Prospecting}, 58(5):861--873, 2010.

\bibitem{nakata2016reverse}
Nori Nakata and Gregory~C Beroza.
\newblock Reverse time migration for microseismic sources using the geometric mean as an imaging condition.
\newblock {\em Geophysics}, 81(2):KS51--KS60, 2016.

\bibitem{wang2017time}
H~Wang and T~Alkhalifah.
\newblock Time reversal migration for passive sources using a maximum variance imaging condition.
\newblock In {\em 79th EAGE Conference and Exhibition 2017}, volume 2017, pages 1--5. European Association of Geoscientists \& Engineers, 2017.

\bibitem{lyu2020iterative}
Bin Lyu and Nori Nakata.
\newblock Iterative passive-source location estimation and velocity inversion using geometric-mean reverse-time migration and full-waveform inversion.
\newblock {\em Geophysical Journal International}, 223(3):1935--1947, 2020.

\bibitem{chen2015directly}
Ting Chen and Lianjie Huang.
\newblock Directly imaging steeply-dipping fault zones in geothermal fields with multicomponent seismic data.
\newblock {\em Geothermics}, 57:238--245, 2015.

\bibitem{bergen2019preface}
Karianne~J Bergen, Ting Chen, and Zefeng Li.
\newblock Preface to the focus section on machine learning in seismology.
\newblock {\em Seismological Research Letters}, 90(2A):477--480, 2019.

\bibitem{wang2018successful}
Zhen Wang, Haibin Di, Muhammad~Amir Shafiq, Yazeed Alaudah, and Ghassan AlRegib.
\newblock Successful leveraging of image processing and machine learning in seismic structural interpretation: A review.
\newblock {\em The Leading Edge}, 37(6):451--461, 2018.

\bibitem{xiong2018seismic}
Wei Xiong, Xu~Ji, Yue Ma, Yuxiang Wang, Nasher~M AlBinHassan, Mustafa~N Ali, and Yi~Luo.
\newblock Seismic fault detection with convolutional neural network.
\newblock {\em Geophysics}, 83(5):O97--O103, 2018.

\bibitem{zheng2019applications}
York Zheng, Qie Zhang, Anar Yusifov, and Yunzhi Shi.
\newblock Applications of supervised deep learning for seismic interpretation and inversion.
\newblock {\em The Leading Edge}, 38(7):526--533, 2019.

\bibitem{hu2020seismic}
Guang Hu, Zhengwang Hu, Jiangping Liu, Fei Cheng, and Daicheng Peng.
\newblock Seismic fault interpretation using deep learning-based semantic segmentation method.
\newblock {\em IEEE Geoscience and Remote Sensing Letters}, 19:1--5, 2020.

\bibitem{ma2022small}
Xiaofei Ma and Ting Chen.
\newblock Small seismic events in oklahoma detected and located by machine learning--based models.
\newblock {\em Bulletin of the Seismological Society of America}, 112(6):2859--2869, 2022.

\bibitem{ANIKIEV2023104371}
Denis Anikiev, Claire Birnie, Umair bin Waheed, Tariq Alkhalifah, Chen Gu, Dirk~J. Verschuur, and Leo Eisner.
\newblock Machine learning in microseismic monitoring.
\newblock {\em Earth-Science Reviews}, 239:104371, 2023.

\bibitem{dando2021relocating}
BDE Dando, BP~Goertz-Allmann, D~K{\"u}hn, N~Langet, AM~Dichiarante, and V~Oye.
\newblock Relocating microseismicity from downhole monitoring of the decatur ccs site using a modified double-difference algorithm.
\newblock {\em Geophysical Journal International}, 227(2):1094--1122, 2021.

\bibitem{alkhalifah2022mlreal}
Tariq Alkhalifah, Hanchen Wang, and Oleg Ovcharenko.
\newblock Mlreal: Bridging the gap between training on synthetic data and real data applications in machine learning.
\newblock {\em Artificial Intelligence in Geosciences}, 3:101--114, 2022.

\bibitem{bauer2016overview}
Robert~A Bauer, Michael Carney, and Robert~J Finley.
\newblock Overview of microseismic response to co2 injection into the mt. simon saline reservoir at the illinois basin-decatur project.
\newblock {\em International Journal of Greenhouse Gas Control}, 54:378--388, 2016.

\bibitem{kaven2014seismic}
J~Ole Kaven, Stephen~H Hickman, Arthur~F McGarr, Steven Walter, and William~L Ellsworth.
\newblock Seismic monitoring at the decatur, il, co2 sequestration demonstration site.
\newblock {\em Energy Procedia}, 63:4264--4272, 2014.

\bibitem{qin2023microseismic}
Yan Qin, Jiaxuan Li, Lianjie Huang, Kai Gao, David Li, Ting Chen, Tom Bratton, George El-kaseeh, William Ampomah, Titus Ispirescu, et~al.
\newblock Microseismic monitoring at the farnsworth co2-eor field.
\newblock {\em Energies}, 16(10):4177, 2023.

\bibitem{qin2022influence}
Yan Qin, Xiaowei Chen, Ting Chen, and Rachel~E Abercrombie.
\newblock Influence of fault architecture on induced earthquake sequence evolution revealed by high-resolution focal mechanism solutions.
\newblock {\em Journal of Geophysical Research: Solid Earth}, 127(11):e2022JB025040, 2022.

\bibitem{langet2020}
N.~Langet, B.~Goertz-Allmann, V.~Oye, R.~A. Bauer, S.~Williams-Stroud, A.M. Dichiarante, and S.E. Greenberg.
\newblock Joint focal mechanism inversion using downhole and surface monitoring at the decatur, illinois, co2 injection site.
\newblock {\em Bull. seism. Soc. Am.}, 110(5):2168--2187, 2020.

\bibitem{albaric2014}
J.~Albaric, V.~Oye, N.~Langet, M.~Hasting, I.~Lecomte, K.~Iranpour, M.~Messeiller, and P.~Reid.
\newblock Monitoring of induced seismicity during the first geothermal reservoir stimulation at paralana, australia.
\newblock {\em Geothermics}, 52:120--131, 2014.

\bibitem{srivastava2014dropout}
Nitish Srivastava, Geoffrey Hinton, Alex Krizhevsky, Ilya Sutskever, and Ruslan Salakhutdinov.
\newblock Dropout: a simple way to prevent neural networks from overfitting.
\newblock {\em The journal of machine learning research}, 15(1):1929--1958, 2014.

\bibitem{6639346}
George~E. Dahl, Tara~N. Sainath, and Geoffrey~E. Hinton.
\newblock Improving deep neural networks for lvcsr using rectified linear units and dropout.
\newblock In {\em 2013 IEEE International Conference on Acoustics, Speech and Signal Processing}, pages 8609--8613, 2013.

\bibitem{fomel2013madagascar}
Sergey Fomel, Paul Sava, Ioan Vlad, Yang Liu, and Vladimir Bashkardin.
\newblock Madagascar: Open-source software project for multidimensional data analysis and reproducible computational experiments.
\newblock {\em Journal of Open Research Software}, 1(1):e8--e8, 2013.

\bibitem{simonyan2015deep}
Karen Simonyan and Andrew Zisserman.
\newblock Very deep convolutional networks for large-scale image recognition, 2015.

\bibitem{wang2021data}
Hanchen Wang, Tariq Alkhalifah, Umair bin Waheed, and Claire Birnie.
\newblock Data-driven microseismic event localization: An application to the oklahoma arkoma basin hydraulic fracturing data.
\newblock {\em IEEE Transactions on Geoscience and Remote Sensing}, 60:1--12, 2021.

\bibitem{wang2021direct}
Hanchen Wang and Tariq Alkhalifah.
\newblock Direct microseismic event location and characterization from passive seismic data using convolutional neural networks.
\newblock {\em Geophysics}, 86(6):KS109--KS121, 2021.

\bibitem{long2015fully}
Jonathan Long, Evan Shelhamer, and Trevor Darrell.
\newblock Fully convolutional networks for semantic segmentation.
\newblock In {\em Proceedings of the IEEE conference on computer vision and pattern recognition}, pages 3431--3440, 2015.

\bibitem{newell2016stacked}
Alejandro Newell, Kaiyu Yang, and Jia Deng.
\newblock Stacked hourglass networks for human pose estimation.
\newblock In {\em European conference on computer vision}, pages 483--499. Springer, 2016.

\bibitem{cao2017realtime}
Zhe Cao, Tomas Simon, Shih-En Wei, and Yaser Sheikh.
\newblock Realtime multi-person 2d pose estimation using part affinity fields.
\newblock In {\em Proceedings of the IEEE Conference on Computer Vision and Pattern Recognition}, pages 7291--7299, 2017.

\bibitem{zhang2016exploiting}
Jianming Zhang and Stan Sclaroff.
\newblock Exploiting surroundedness for saliency detection: a boolean map approach.
\newblock {\em IEEE transactions on pattern analysis and machine intelligence}, 38(5):889--902, 2016.

\bibitem{zhou2016learning}
Bolei Zhou, Aditya Khosla, Aude Lapedriza, Aude Oliva, and Antonio Torralba.
\newblock Learning deep features for discriminative localization.
\newblock In {\em Proceedings of the IEEE conference on computer vision and pattern recognition}, pages 2921--2929, 2016.

\bibitem{springenberg2014striving}
Jost~Tobias Springenberg, Alexey Dosovitskiy, Thomas Brox, and Martin Riedmiller.
\newblock Striving for simplicity: The all convolutional net.
\newblock {\em arXiv preprint arXiv:1412.6806}, 2014.

\bibitem{selvaraju2017grad}
Ramprasaath~R Selvaraju, Michael Cogswell, Abhishek Das, Ramakrishna Vedantam, Devi Parikh, and Dhruv Batra.
\newblock Grad-cam: Visual explanations from deep networks via gradient-based localization.
\newblock In {\em Proceedings of the IEEE International Conference on Computer Vision}, pages 618--626, 2017.

\bibitem{kingma2017adam}
Diederik~P. Kingma and Jimmy Ba.
\newblock Adam: A method for stochastic optimization, 2017.

\bibitem{kendall2017uncertainties}
Alex Kendall and Yarin Gal.
\newblock What uncertainties do we need in bayesian deep learning for computer vision?
\newblock {\em Advances in neural information processing systems}, 30, 2017.

\bibitem{williams2020analysis}
Sherilyn Williams-Stroud, Robert Bauer, Hannes Leetaru, Volker Oye, Frantisek Stanek, Sallie Greenberg, and Nadege Langet.
\newblock Analysis of microseismicity and reactivated fault size to assess the potential for felt events by co 2 injection in the illinois basin.
\newblock {\em Bulletin of the Seismological Society of America}, 110(5):2188--2204, 2020.

\bibitem{skoumal2019characterizing}
Robert~J Skoumal, J~Ole Kaven, and Jacob~I Walter.
\newblock Characterizing seismogenic fault structures in oklahoma using a relocated template-matched catalog.
\newblock {\em Seismological Research Letters}, 90(4):1535--1543, 2019.

\bibitem{illinois_netl}
NETL.
\newblock Illinois state geological survey (isgs), illinois basin - decatur project (ibdp) seismic data, july 7, 2021. midwest geological sequestration consortium (mgsc) phase iii data sets. doe cooperative agreement no. de-fc26-05nt42588 [dataset], 2021.

\end{thebibliography}

\end{document}